\documentclass[
 reprint,
superscriptaddress,
 amsmath,amssymb,
 aps,
pra,
]{revtex4-2}

\usepackage{graphicx}% Include figure files
\usepackage{dcolumn}% Align table columns on decimal point
\usepackage{bm}% bold math

\DeclareMathAlphabet{\pazocal}{OMS}{zplm}{m}{n}
\DeclareMathOperator{\tr}{tr}
\usepackage[colorlinks=true, linktocpage=true]{hyperref}
\usepackage{blindtext}
\hypersetup{colorlinks=true, linktoc=all, citecolor=blue,linkcolor=blue,urlcolor=blue}
\usepackage{hyperref}
\usepackage{orcidlink}

\begin{document}

\preprint{APS/123-QED}
%\title{microscopic derivation of Open Quantum Brownian Motion: adiabatic elimination approach} 

\title{Adiabatic elimination and Wigner function approach in microscopic derivation of Open Quantum Brownian Motion}

\author{Ayanda Zungu\orcidlink{0000-0001-5972-9179}}
\email[]{ayanda.zungu@nwu.ac.za}
%\affiliation{School of Chemistry and Physics, University of KwaZulu-Natal, Durban 4001, South Africa}
\affiliation{Discipline of Physics, School of Agriculture and Science, University of KwaZulu-Natal, Durban 4001, South Africa}
\affiliation{Centre for Space Research, North-West University, Mahikeng 2745, South Africa}
\affiliation{National Institute for Theoretical and Computational Sciences (NITheCS), North-West University, Mahikeng, South Africa}

\author{Ilya Sinayskiy\orcidlink{0000-0002-3040-0051
}}
\email[]{sinayskiy@ukzn.ac.za}
%\affiliation{School of Chemistry and Physics, University of KwaZulu-Natal, Durban 4001, South Africa}
\affiliation{Discipline of Physics, School of Agriculture and Science, University of KwaZulu-Natal, Durban 4001, South Africa}

\affiliation{National Institute for Theoretical and Computational Sciences (NITheCS), University of KwaZulu-Natal, Durban, South Africa}

\author{Francesco Petruccione\orcidlink{0000-0002-8604-0913
}}
%\email[]{sinayskiy@ukzn.ac.za}
\affiliation{National Institute for Theoretical and Computational Sciences (NITheCS), Stellenbosch, South Africa}

\affiliation{School of Data Science and Computational Thinking and Department of Physics, Stellenbosch University, Stellenbosch 7604, South Africa}
 
\date{\today}
%%%%%%%%%%%%%%%%%%%%%%%%%%%%%%%%%%%%%%%%%%%%%%%%%%%%%%%%%%%%%
\begin{abstract}
%%%%%%%%%%%%%%%%%%%%%%%%%%%%%%%%%%%%%%%%%%%%%%%%%%%%%%%%%%%%%
Open Quantum Brownian Motion (OQBM) is a new class of quantum Brownian motion in which the dynamics of the Brownian particle depend not only on interactions with a thermal environment but also on the state of its internal degrees of freedom. For an Ohmic bath spectral density with a Lorentz--Drude cutoff frequency at a high-temperature limit, we derive the Born--Markov master equation for the reduced density matrix of an open Brownian particle in a harmonic potential. The resulting master equation is written in phase-space representation using the Wigner function, and due to the separation of associated timescales in the high-damping limit, we perform adiabatic elimination of the momentum variable to obtain OQBM. We numerically solve the derived master equation for the reduced density matrix of the OQBM, using both Gaussian and non-Gaussian initial distributions. In each case, the OQBM dynamics converge to several Gaussian distributions. To gain physical insight into the studied system, we also plotted the dynamics of the off-diagonal element of the open quantum Brownian particle and found damped coherent oscillations. Finally, we investigated the time-dependent variance in the position of the OQBM walker and observed initial diffusive spreading, followed by a crossover to a finite steady-state value.

%Finally, we investigated the time-dependent variance in the position of the OQBM walker and observed a transition between ballistic and diffusive behavior.
\end{abstract}

\maketitle
%%%%%%%%%%%%%%%%%%%%%%%%%%%%%%%%%%%%%%%%%%%%%%%%%%%%%%%%%%%%%
\section{\label{sec:level1} Introduction}
%%%%%%%%%%%%%%%%%%%%%%%%%%%%%%%%%%%%%%%%%%%%%%%%%%%%%%%%%%%%%
A physical system interacting with its surroundings is called an open system. Such interactions are inevitable, as isolated physical systems are an idealization. The coupling of the quantum system to the environment causes dissipation, thermalization, and decoherence~\cite{breuer2002theory}. 
Usually, these processes lead to the destruction of the quantumness of the system, thereby reducing the computational power of quantum computers by reducing the fidelity of quantum gates and introducing errors in computations. Such effects should be minimized or controlled in quantum computation, communication, and simulation. Therefore, techniques for simulating the dynamics of open quantum systems are vital for developing quantum technologies.

The Lindblad master equation~\cite{lindblad1976generators,gorini1976completely} governs the non-reversible evolution of various system-bath coupling regimes, typically for systems weakly coupled to a Markovian bath. To investigate the effects of dissipation and decoherence in unitary quantum walks (UQWs)~\cite{aharonov1993quantum,kempe2003quantum}, which have been used as a basic tool for designing effective quantum algorithms and universal quantum computation~\cite{PhysRevLett.102.180501,venegas2012quantum, PhysRevA.81.042330,chawla2023multi}, a new class of non-unitary quantum walks called open quantum walks (OQWs) were introduced to consider the dynamic behavior of open quantum systems~\cite{ATTAL20121545,attal2012open, Sinayskiy_2012}. 

OQWs are fundamentally distinct from UQWs, exhibiting qualitatively different properties.
On graphs or lattices, OQWs are expressed as quantum Markov chains and are represented mathematically by completely positive trace-preserving (CPTP) maps~\cite{breuer2002theory,kraus1983states}. The CPTP maps correspond to dissipative processes that drive transitions between the nodes. Unlike UQWs, which use quantum interference effects~\cite{aharonov1993quantum,kempe2003quantum,venegas2012quantum}, in OQWs, the interaction with the environment strictly drives the transitions between the nodes. Accordingly, the environment significantly impacts how OQWs evolve. 
OQWs use density matrices rather than a pure state, and they admit central limit theorems~\cite{konno2013limit,sadowski2016central,attal2015central}, which is a crucial distinction between UQWs and OQWs. 
OQWs exhibit rich dynamics, making them a fascinating field of study for quantum computing and quantum information. 
For example, OQWs naturally begin as quantum walks and, over long time limits, transform into classical random walks; for significant times, the position probability distribution of OQWs converges to a Gaussian distribution or a mixture of Gaussian distributions, consistent with the OQWs central limit theorem~\cite{konno2013limit}.

Moreover, it has been suggested that OQWs can generate complex quantum states and perform dissipative quantum computation~\cite{ATTAL20121545,attal2012open, Sinayskiy_2012,sinayskiy2012efficiency}. In addition, the
discrete-time OQWs have been generalized to continuous-time OQWs~\cite{pellegrini2014continuous}. The complete description of the OQW framework can be found in~\cite{ATTAL20121545,attal2012open, Sinayskiy_2012}, and a recent article~\cite{sinayskiy2019open} reviews progress on this subject. More crucially,~\cite{sinayskiy2014quantum} suggested a quantum optics implementation of OQWs, and then showed that OQWs can be derived from the microscopic system-bath model~\cite{sinayskiy2013microscopic, PhysRevA.92.032105}. 

Bauer~\textit{et al.}~\cite{bauer2013open,bauer2014open} introduced open quantum Brownian motion (OQBM) as a scaling limit of OQWs, which represents a new type of quantum Brownian motion with one additional quantum internal degree of freedom, and the microscopic derivation of OQBM for the case of a free Brownian particle and decoherent interaction with an environment has been suggested~\cite{sinayskiy2015microscopicbrown,sinayskiy2017steady}. The OQBM considered in Refs.~\cite{sinayskiy2015microscopicbrown,sinayskiy2017steady} was subsequently analyzed by~\cite{Carlos2025}.
However, the microscopic derivation in a generic dissipative case is still
missing. 
In this paper, we derive the OQBM for a Brownian particle in a harmonic potential that interacts dissipatively with a thermal bath, using an adiabatic elimination method. 
Although this method has previously been used in the literature~\cite{SKINNER1979561,van1985elimination,kramers1940brownian,gardiner1985handbook,Gardiner1984}, it has never been used before to derive OQBM. 

The model under consideration consists of a Brownian particle with a single quantum internal degree of freedom trapped in a harmonic potential. The Brownian particle is weakly coupled to a thermal bath that is made up of a large number of bosonic harmonic oscillators. The internal degree of freedom is modelled as a two-level system, and the position operator describes the external degrees of freedom.
Starting from the Hamiltonian of the quantum Brownian particle with a single internal degree of freedom, the Hamiltonian of the bath, and the Hamiltonian of the system-bath interaction, we derive the Born--Markov master equation for the reduced density matrix. 

The resulting master equation is written in phase-space representation using the Wigner function. 
In the high-damping limit, we assume that the Brownian particle's momentum dissipates quickly to the steady state while the position variable evolves more slowly. Essentially, we derive a variant of the Quantum Smoluchowski equation~\cite{Ankerhold2001,PhysRevE.81.021107}. The resulting timescale separation between momentum and position allows us to perform adiabatic elimination of the momentum variable, thereby obtaining the OQBM. 
Using this method, we derive a master equation that includes diffusive, dissipative, and ``decision-making'' terms, and it has the same structure as the form originally suggested by Bauer~\textit{et al.}~\cite{bauer2013open,bauer2014open} and demonstrated by~\cite{sinayskiy2015microscopicbrown,sinayskiy2017steady}.
The master equation describing OQBM is a typical example of a hybrid quantum-classical master equation~\cite{Diósi_2014,PhysRevA.107.062206}, and such equations are becoming prominent in the literature, especially in theories related to gravity~\cite{layton2024healthier,layton2024classical,oppenheim2023postquantum,oppenheim2022two,halliwell1998effective,tilloy2024general}.

The structure of the paper is as follows. In Sect.~\ref{part_1}, we start from the microscopic Hamiltonian and derive the Born--Markov master equation for the Brownian particle with a single quantum internal degree of freedom. In Sect.~\ref{part_2}, we present systematic adiabatic elimination of the momentum variable and obtain a master equation that defines OQBM. Sect.~\ref{part_3} contains numerical examples of OQBM dynamics and discussions. 
The $n$th moments of the OQBM walker's position distribution are derived in Sect.~\ref{part_4} using the OQBM master equation~(\ref{finmee}) and are solved numerically for various system-bath parameters. 
In Sect.~\ref{part_5}, we summarize the results of the paper.

%%%%%%%%%%%%%%%%%%%%%%%%%%%%%%%%%%%%%%%%%%%%%%%%%%%%%%%%%%%%%
\section{Microscopic derivation\label{part_1}}
%%%%%%%%%%%%%%%%%%%%%%%%%%%%%%%%%%%%%%%%%%%%%%%%%%%%%%%%%%%%%
This section presents the microscopic derivation of the Born--Markov master equation for a quantum Brownian particle with a single internal degree of freedom. The position operator $\hat{x}$ describes the external degree of freedom of a Brownian particle, and the internal degree of freedom is described by a two-level system (2LS). 
The model for the dynamics of this dissipative quantum system is obtained by weakly coupling the system of interest to a Markovian bath~\cite{breuer2002theory}. 
The following Hamiltonian defines the model
\begin{equation}\label{hbr}
    \hat{H} = \hat{H}_S+\hat{H}_B+\hat{H}_{SB}+\hat{H}_c,
\end{equation}
\noindent
%where the system, bath, and the system-bath interaction Hamiltonians are respectively given by
where the system, bath, system-bath interaction Hamiltonians, and the counterterm are, respectively, given by
\begin{align}
         \hat{H}_S &= \frac{\hat{p}^2}{2m}+\frac{m\omega^2\hat{x}^2}{2}+\frac{\hbar\Omega}{2}\hat{\sigma}_z,\label{hbrwn}\\
        \hat{H}_B &= \sum_n \frac{\hat{p}_n^2}{2m_n}+\frac{m_n\omega_n^2\hat{x}_n^2}{2},\label{hbr_bath}\\
         \hat{H}_{SB} &= \sum_n g_n\hat{x}_n\hat{x}+C_n\hat{x}_n\hat{\sigma}_x,\label{hbr_interact}\\
         \hat{H}_c &=\hat{x}^2\sum_n\frac{g_n^2}{2m_n\omega_n^2}.\label{counter_t}
\end{align}
\noindent 
Here, $m$ is the mass of the Brownian particle, $\omega$ is the frequency of the harmonic potential trapping it, and $\hat{p}$ is the momentum operator.
The bath is modelled by $n$th quantum harmonic oscillators, described by $m_n$, $\hat{x}_n$, $\omega_n$, $\hat{p}_n$, which denote the mass, coordinates, natural frequency, and momentum, respectively. The operators $\hat{x}_n$ and $\hat{p}_n$ satisfy the usual commutation relation $[\hat{x}_n,\hat{p}_m]=i\hbar\delta_{n,m}$.

The first two terms in Eq.~(\ref{hbrwn}) describe the Hamiltonian of a single quantum harmonic oscillator, and the last term corresponds to the Hamiltonian of the free 2LS, with $\Omega$ denoting the transition frequency, $\hat{\sigma}_k$ ($k=x,y,z$) are the Pauli matrices. 
The open Brownian particle is coupled linearly to each oscillator, with the bath-particle coupling constant denoted by $g_n$, while $C_n$ denotes the coupling constant to the internal degree of freedom. 
The last term~(\ref{counter_t}) denotes an additional counterterm that compensates the bath-induced potential renormalization
and acts only in the Hilbert space $\mathcal{H}_S$ of the system~\cite{breuer2002theory}.

To derive the Born--Markov master equation, we start from the microscopic Hamiltonian~(\ref{hbr}) and follow the traditional techniques of the theory of open quantum systems~\cite {breuer2002theory} and derive the reduced dynamics. 
%To derive the Born--Markov master equation, we start from the microscopic Hamiltonian~(\ref{hbr}) and follow the traditional techniques of the theory of open quantum systems~\cite{breuer2002theory}, assuming $\hbar\gamma \ll \min\{\hbar\Lambda,2\pi k_B T\}$, where $\gamma$ is the damping rate, $\Lambda$ is the bath cutoff frequency, and $T$ is the bath temperature.
The reduced density matrix $\hat{\rho}_S (t)$ corresponding to the system of interest is obtained from the density matrix of the total system $\hat{\rho}_{SB} (t)$ by taking the partial trace over the bath degrees of freedom, i.e., $\hat{\rho}_S (t) = \tr_B\bigl[\hat{\rho}_{SB}(t)\bigl].$

We assume the system and bath are initially ($t=0$) uncorrelated, i.e., the initial density matrix is given by the tensor product, $\hat{\rho}_{SB}(0)=\hat{\rho}_S(0)\otimes\hat{\rho}_B(0)$.
We then assume that the system and the bath are weakly coupled (Born approximation), which means that the influence of the system on the bath is negligible and the total system remains roughly uncorrelated for all times, i.e., $\hat{\rho}_{SB}(t)\approx \hat{\rho}_S(t) \otimes\hat{\rho}_B(0).$
The bath is assumed to be in thermal equilibrium at temperature $T$, i.e., its density matrix $\hat{\rho}_B(0)$ is given by
\begin{equation}
    \hat{\rho}_B(0) = \frac{1}{\mathcal{Z}}e^{-\beta \hat{H}_B}, \hspace{4mm} \text{where} \hspace{4mm} \mathcal{Z}=\tr_B \Bigl[ e^{-\beta \hat{H}_B}\Bigl].
\end{equation}
\noindent
Here, $\mathcal{Z}$ denotes the partition function, $\beta=({k_B T})^{-1}$ and $k_B$ is the Boltzmann constant. 
The master equation for the reduced dynamics $\hat{\rho}_S(t)$ is obtained by starting from the traditional Born--Markov master equation~\cite{breuer2002theory,schlosshauer2007decoherence}. 
When applied to our system, the Born--Markov master equation becomes
\begin{align}\label{semeq}
    \frac{d}{dt}\hat{\rho}_S (t) =& -\frac{i}{\hbar}\bigl[\hat{H}_S+\hat{H}_c,\hat{\rho}_S\bigl]
    -\frac{1}{\hbar^2}\int_0^\infty d\tau \tr_B\Bigl\{\Bigl[\hat{H}_{SB}(0),\nonumber\\
    &\bigl[\hat{H}_{SB}(-\tau),\hat{\rho}_S(t)\otimes\hat{\rho}_B(0)\bigl]\Bigl]\Bigl\},
\end{align}
\noindent
where $\hat{H}_{SB}(-\tau)$ is the Hamiltonian of the system-bath interaction in the interaction picture, given by
\begin{align}\label{intHb}
    \hat{H}_{SB}&(-\tau)=\sum_n g_n e^{-i\tau(\hat{H}_S+\hat{H}_B)/\hbar}\hat{x}_n\hat{x}e^{i\tau(\hat{H}_S+\hat{H}_B)/\hbar}\nonumber\\
    &+C_ne^{-i\tau(\hat{H}_S+\hat{H}_B)/\hbar}\hat{x}\hat{\sigma}_xe^{i\tau(\hat{H}_S+\hat{H}_B)/\hbar}\nonumber\\
    =&\sum_n g_n\hat{x}_n(-\tau)\hat{x}(-\tau)+C_n\hat{x}_n(-\tau)\hat{\sigma}_x(-\tau).
\end{align}
\noindent
In Eq.~(\ref{intHb}), $\hat{x}(-\tau)$, $\hat{x}_n(-\tau)$, and $\hat{\sigma}_x(-\tau)$ are the standard Heisenberg-picture expressions given respectively by 
\begin{align}\label{interterms}
\hat{x}(-\tau) &= \hat{x} \cos\omega\tau-\frac{\hat{p}}{m\omega}\sin\omega\tau,\nonumber\\
    \hat{x}_n(-\tau) &= \hat{x}_n \cos\omega_n\tau-\frac{\hat{p}_n}{m_n\omega_n}\sin\omega_n\tau,\nonumber\\ 
    \hat{\sigma}_x(-\tau) &=\hat{\sigma}_x\cos\Omega\tau+\hat{\sigma}_y\sin\Omega\tau.
\end{align}
\noindent
By using Eq.~(\ref{semeq}) together with Eq.~(\ref{interterms}) and keeping only the terms with the same indices (terms with different indices are completely uncorrelated and are all equal to zero), one obtains the following master equation
\begin{equation}\label{mastersep}
    \frac{d}{dt}\hat{\rho}_S (t) = \pazocal{L}_{\mathrm{QHO}}\hat{\rho}_S+\pazocal{L}_{\mathrm{2LS}}\hat{\rho}_S+\pazocal{L}_{\mathrm{cross}}\hat{\rho}_S,
\end{equation}
\noindent
where $\pazocal{L}_{\mathrm{QHO}}\hat{\rho}_S$, $\pazocal{L}_{\mathrm{2LS}}\hat{\rho}_S$, and $\pazocal{L}_{\mathrm{cross}}\hat{\rho}_S$ denote the dissipators of the quantum harmonic oscillator, 2LS, and 
the ``cross-term'' dissipator that captures dissipative coupling between external and internal degrees of freedom, which are respectively given by
\begin{align} 
\pazocal{L}_{\mathrm{QHO}}&\hat{\rho}_S=-\frac{i}{\hbar}\bigl[\hat{H}_\mathrm{QHO}+\hat{H}_c,\hat{\rho}_S\bigl]-\frac{1}{\hbar^2}\int_0^\infty d\tau\sum_n |g_n|^2 \nonumber\\
&\tr_B\Bigl\{\Bigl[\hat{x}_n\hat{x},\bigl[ \hat{x}_n(-\tau)\hat{x}(-\tau),\hat{\rho}_S(t)\otimes\hat{\rho}_B(0)\bigl]\Bigl]\Bigl\},\label{brw0}\\
\pazocal{L}_{\mathrm{2LS}}&\hat{\rho}_S=-\frac{i\Omega}{2}\bigl[\hat{\sigma}_z,\hat{\rho}_S\bigl]-\frac{1}{\hbar^2}\int_0^\infty d\tau\sum_n|C_n|^2\nonumber\\
& \tr_B\Bigl\{\Bigl[\hat{x}_n\hat{\sigma}_x,\bigl[ \hat{x}_n(-\tau)\hat{\sigma}_x(-\tau),\hat{\rho}_S(t)\otimes\hat{\rho}_B(0)\bigl]\Bigl]\Bigl\},\label{brw1}\\
\pazocal{L}_{\mathrm{cross}}&\hat{\rho}_S =  -\frac{1}{\hbar^2}\int_0^\infty d\tau\sum_n g_nC_n^* \tr_B\Bigl\{\Bigl[\hat{x}_n\hat{x},\nonumber\\
&\bigl[ \hat{x}_n(-\tau)\hat{\sigma}_x(-\tau),\hat{\rho}_S(t)\otimes\hat{\rho}_B(0)\bigl]\Bigl]\Bigl\}\nonumber\\
  &-\frac{1}{\hbar^2}\int_0^\infty d\tau\sum_n g_nC_n^* \tr_B\Bigl\{\Bigl[\hat{x}_n\hat{\sigma}_x,\nonumber\\
&\bigl[ \hat{x}_n(-\tau)\hat{x}(-\tau),\hat{\rho}_S(t)\otimes\hat{\rho}_B(0)\bigl]\Bigl]\Bigl\},\label{brw2}
\end{align}
\noindent
where $\hat{H}_\mathrm{QHO}=\frac{\hat{p}^2}{2m}+\frac{m\omega^2\hat{x}^2}{2}$. 
The next step is to evaluate the bath self-correlation function $ \mathcal{C}(-\tau)$, given by
\begin{equation}\label{eq:b}
    \mathcal{C}(-\tau) = \sum_{n} |\kappa_n|^2 \bigl\langle \hat{x}_n \hat{x}_n (-\tau)\bigl\rangle_B,
\end{equation}
\noindent
where $\kappa_n\in(g_n, C_n)$. The expression $\langle \cdots \rangle_B$ in Eq.~(\ref{eq:b}) is evaluated to be
\begin{align}
    & \langle \hat{x}_n \hat{x}_n (-\tau)\rangle_B = \frac{\hbar}{2m_n\omega_n}\bigl[(2n(\omega_n)+1)\cos(\omega_n \tau)\nonumber\\
     &-i\sin(\omega_n \tau)\bigl]\nonumber\\
    &=  \frac{\hbar}{2m_n\omega_n}\bigl[\coth(\hbar\beta \omega_n/2)\cos(\omega_n \tau)-i\sin(\omega_n \tau)\bigl],
\end{align}
\noindent
where $n(\omega_n) = [\exp(\hbar \beta\omega_n)-1]^{-1}$ represents the mean bosonic occupation number.
Hence, the bath self-correlation function for the quantum harmonic oscillator is given by
\begin{align}\label{scorr}
     \mathcal{C}(-\tau)=&\sum_n\frac{\hbar |g_n|^2}{2m_n\omega_n}\bigl[\coth(\hbar\beta \omega_n/2)\cos(\omega_n \tau)\nonumber\\
     &-i\sin(\omega_n \tau)\bigl]\nonumber\\
     \equiv&\ \nu (\tau)-i\eta (\tau),
\end{align}
\noindent
where the thermal noise kernel $\nu (\tau)$ is
\begin{align}\label{noise}
    \nu (\tau) &= \sum_n\frac{\hbar |g_n|^2}{2m_n\omega_n}\coth(\hbar\beta \omega_n/2)\cos(\omega_n \tau)\nonumber\\
    &\equiv \hbar \int_0^\infty d\omega \,J(\omega)\coth(\hbar\beta \omega/2)\cos(\omega \tau),
\end{align}
\noindent
and the dissipation kernel $\eta (\tau)$ is
\begin{align}\label{dissi}
    \eta (\tau) = \sum_n\frac{\hbar |g_n|^2}{2m_n\omega_n}\sin(\omega_n \tau)\equiv \hbar \int_0^\infty d\omega\, J(\omega)\sin(\omega \tau).
\end{align}
\noindent
In Eqs.~(\ref{noise})-(\ref{dissi}), we defined a continuous frequency density distribution function instead of the discrete oscillator distribution, i.e., $\sum_n \rightarrow \int d\omega$. The function $J(\omega)$ is the spectral density, and it 
arises from the extra information supplied by the microscopics of the bath constituents
\begin{equation}
    J(\omega) =\sum_n\frac{|\kappa_n|^2}{2m_n\omega_n}\delta(\omega-\omega_n).
\end{equation}
\noindent
%In the next step, we consider the most fundamental example of an open quantum system: the quantum Brownian motion. 
In the next step, we consider a paradigmatic example of an open quantum system, namely quantum Brownian motion~\cite{breuer2002theory}. Using the self-correlation function~(\ref{scorr}), we can now write the Liouvillian superoperator~(\ref{brw0}) in the simple form as
\begin{align}\label{qhomast}
&\pazocal{L}_{\mathrm{QHO}}\hat{\rho}_S=-\frac{i}{\hbar}\bigl[\hat{H}_\mathrm{QHO}+\hat{H}_c,\hat{\rho}_S\bigl] -\frac{D_x}{\hbar^2}\bigl[\hat{x},[\hat{x},\hat{\rho}_S]\bigl]\nonumber\\
&+\frac{D_p}{\hbar^2m\omega}\bigl[\hat{x},[\hat{p},\hat{\rho}_S]\bigl]+\frac{iC_x}{\hbar^2}\bigl[\hat{x},\{\hat{x},\hat{\rho}_S\}\bigl]-\frac{iC_p}{\hbar^2m\omega} \bigl[\hat{x},\{\hat{p},\hat{\rho}_S\}\bigl],
\end{align}
\noindent
with $\{\cdot,\cdot\}$ indicating the anti-commutator. The coefficients appearing in Eq.~(\ref{qhomast}) are given by
 \begin{align}\label{coeff1}
D_x &= \int_0^\infty d\tau \,\nu (\tau)\cos\omega \tau,\hspace{1mm}
C_x = \int_0^\infty d\tau\, \eta (\tau)\cos\omega \tau,\nonumber \\
 D_p &= \int_0^\infty d\tau\, \nu (\tau)\sin\omega \tau,\hspace{1mm}
C_p = \int_0^\infty d\tau\, \eta (\tau)\sin\omega \tau,
     \end{align}
\noindent
and can be evaluated explicitly for a specific spectral density $J(\tilde{w})$. In this paper, we adopt an Ohmic spectral density  $J(\tilde{w}) \propto \tilde{\omega}$ with a Lorentz--Drude cutoff of the following form~\cite{breuer2002theory}:
\begin{equation}\label{spectra}
    J(\tilde{\omega}) = \frac{2m\gamma}{\pi}\tilde{\omega}\frac{\Lambda^2}{\Lambda^2+\tilde{\omega}^2}.
\end{equation}
\noindent
Here, the constant $\gamma$ is the frequency-independent damping coefficient, and $\Lambda$ is the high-frequency cutoff. The Born--Markov approximation requires
$\hbar\gamma \ll \min\{\hbar\Lambda,2\pi k_B T\}$~\cite{breuer2002theory}.
The spectral density~(\ref{spectra}) allows us to explicitly 
compute the coefficients $D_x$, $C_x$, $D_p$, and $C_p$. As a first step, we expand $\coth(\cdot)$ using the Matsubara representation~\cite{breuer2002theory}:
\begin{equation}\label{mastu}
\coth\biggl({\frac{\hbar\tilde{\omega}}{2k_B T}}\biggl) = \frac{2k_B T}{\hbar\tilde{\omega}}+\frac{4k_B T}{\hbar\tilde{\omega}}\sum_{n=1}^\infty\frac{1}{1+(\nu_n/\tilde{\omega})^2},
\end{equation}
\noindent
where the $\nu_n=2\pi n k_B T / \hbar$ are known as the Matsubara frequencies. 
In the limit of high temperatures $k_B T \gg \hbar \tilde{\omega}$, $\coth\bigl(\hbar\tilde{\omega}/2k_B T\bigl) \approx 2k_BT/\hbar\tilde{\omega}$, the thermal noise kernel~(\ref{noise}), and the dissipation kernel~(\ref{dissi}) are evaluated analytically to be
\begin{align}\label{kerns}
   \nu (\tau) = \ & 2m\gamma k_B T   \Lambda e^{-\Lambda|\tau|}, \nonumber\\
   \eta (\tau) = \ & m\gamma \hbar \Lambda^2 \text{sign}(\tau) e^{-\Lambda|\tau|}.
\end{align}
\noindent
Using Eq.~(\ref{kerns}), it is straightforward to show that the coefficients~(\ref{coeff1}) become
\begin{align}\label{coeff3}
D_x &=2m\gamma k_B T \biggl(\frac{\Lambda^2}{\Lambda^2+\omega^2}\biggl), \hspace{2mm}
C_x = m\gamma \hbar \biggl( \frac{\Lambda^3}{\Lambda^2+\omega^2}\biggl),\nonumber \\
 D_p &= 2m\gamma k_B T\Lambda\biggl( \frac{\omega}{\Lambda^2+\omega^2}\biggl),\hspace{2mm}
C_p =  m\gamma \hbar \Lambda^2 \biggl(\frac{\omega}{\Lambda^2+\omega^2}\biggl).
\end{align}
\noindent
Again, in the high-temperature limit and large-cutoff limit $k_B T\gg \hbar\Lambda \gg \hbar\omega$, Eq.~(\ref{coeff3}) reduces to
\begin{align}\label{coeff4}
D_x &\approx 2m\gamma k_B T, \hspace{7mm} C_x \approx m\gamma \hbar \Lambda,\nonumber \\
 D_p &\approx 2m\gamma k_B T\frac{\omega}{\Lambda},\hspace{4mm} C_p \approx  m\gamma \hbar \omega.
\end{align}
\noindent
%Inserting the above expressions~(\ref{coeff4}) into Eq.~(\ref{qhomast}) leads to
Inserting the expressions in Eqs.~(\ref{coeff4}) into Eq.~(\ref{qhomast}) leads to
\begin{align}\label{qhomastf}
&\pazocal{L}_{\mathrm{QHO}}\hat{\rho}_S= -\frac{i}{\hbar}\bigl[\hat{H}_\mathrm{QHO}+\hat{H}_c,\hat{\rho}_S\bigl] -\frac{2m\gamma k_B T}{\hbar^2}\bigl[\hat{x},[\hat{x},\hat{\rho}_S]\bigl]\nonumber\\
&+\frac{2\gamma k_B T}{\hbar^2\Lambda}\bigl[\hat{x},[\hat{p},\hat{\rho}_S]\bigl]+\frac{im\gamma\Lambda}{\hbar}\bigl[\hat{x},\{\hat{x},\hat{\rho}_S\}\bigl]-\frac{i\gamma}{\hbar} \bigl[\hat{x},\{\hat{p},\hat{\rho}_S\}\bigl].
\end{align}
\noindent
The third term on the right-hand side of Eq.~(\ref{qhomastf}) may be neglected because the momentum is of the order of $\hat{p} \sim m \omega \hat{x}$ and it scales as $\omega/\Lambda$, which by assumption is very small. 
The fourth term on the right-hand side of Eq.~(\ref{qhomastf}), $\bigl[\hat{x},\{\hat{x},\hat{\rho}_S\}\bigl] \ = [\hat{x}^2, \hat{\rho}_S]$, is absorbed by the unitary dynamics term and cancels the contribution arising from the counterterm Hamiltonian~(\ref{counter_t}).
Finally, we arrive at the Caldeira--Leggett type superoperator~\cite{caldeira1983path,CALDEIRA1983374}:
\begin{align}\label{qhomastfn}
\pazocal{L}_{\mathrm{QHO}}\hat{\rho}_S=& -\frac{i}{\hbar}\bigl[\hat{H}_\mathrm{QHO},\hat{\rho}_S\bigl] -\frac{2m\gamma k_B T}{\hbar^2}\bigl[\hat{x},[\hat{x},\hat{\rho}_S]\bigl] \nonumber\\
&-\frac{i\gamma}{\hbar} \bigl[\hat{x},\{\hat{p},\hat{\rho}_S\}\bigl].
\end{align}
\noindent
For the subsequent moment analysis, it is more convenient to redefine the operators $\hat{x}$ and $\hat{p}$ in Eq.~(\ref{qhomastfn}) as dimensionless operators by multiplying them by $x_0$ and $p_0$, respectively, to obtain
\begin{align}\label{qhomastfnf}
\pazocal{L}_{\mathrm{QHO}}\hat{\rho}_S=& -\frac{i}{\hbar}\bigl[\hat{H}_\mathrm{QHO},\hat{\rho}_S\bigl] -\frac{2m\gamma k_B T}{\hbar^2}x_0^2\bigl[\hat{x},[\hat{x},\hat{\rho}_S]\bigl] \nonumber\\
&-\frac{i\gamma}{\hbar}p_0x_0 \bigl[\hat{x},\{\hat{p},\hat{\rho}_S\}\bigl],
\end{align}
where $x_0$, $p_0$, and $\hat{H}_\mathrm{QHO}$ are 
\begin{align}
&   x_0 = \sqrt{\frac{\hbar}{2m\omega}}, \hspace{4mm} p_0 = \sqrt{\frac{m\hbar \omega}{2}},\nonumber\\
&\hat{H}_\mathrm{QHO}=\frac{\hat{p}^2}{2m}p_0^2+\frac{m\omega^2\hat{x}^2}{2}x_0^2.
\end{align}
\noindent
The three terms in Eq.~(\ref{qhomastfnf}) have a typical physical interpretation. The first term
on the right-hand side describes the free coherent dynamics. The second term represents thermal fluctuations and is proportional to the temperature, a crucial feature of the theoretical formulation of decoherence. The final term, proportional to the damping coefficient $\gamma$, is the dissipative term. 
Equation~(\ref{qhomastfnf}) describes the reduced dynamics of a quantum harmonic oscillator, which is linearly and weakly coupled to a thermal bath of $n$th harmonic oscillators.

It is well known that master equations such as Eq.~(\ref{qhomastfnf}) violate the positivity constraint of the density matrix~\cite{L.Diósi_1993, homa2019positivity,PhysRevLett.84.1374,PhysRevA.94.042123}, which can often lead to unphysical outcomes. %However, in this work, as is usually done for such types of equations, we are going to consider initial conditions and the evaluation times, which do not lead to unphysical results.
However, in this work, following standard practice for such equations, we consider initial conditions and evaluation times that do not lead to unphysical results.
Here, it is also worth mentioning that we are interested in the classical limit of Eq.~(\ref{qhomastfnf}), where the momentum of the Brownian particle dissipates very fast. This limit will be taken from the phase-space representation of the reduced density matrix.

We now consider the 2LS term~(\ref{brw1}). Noting that $\hat{\sigma}_x(-\tau)$ is
\begin{equation}\label{transtau}
    \hat{\sigma}_x(-\tau)=\hat{\sigma}_+e^{-i\Omega\tau}+\hat{\sigma}_-e^{i\Omega\tau},
\end{equation}
\noindent
Eq.~(\ref{brw1}) can be written in the simplest form as
    \begin{align}\label{2lvmast}
&\pazocal{L}_{\mathrm{2LS}}\hat{\rho}_S=-\frac{i\Omega}{2}[\hat{\sigma}_z,\hat{\rho}_S]+D_{xx}\Bigl(\hat{\sigma}_-\hat{\rho}_S\hat{\sigma}_+-\hat{\sigma}_+\hat{\sigma}_-\hat{\rho}_S\Bigl)\nonumber\\
&+C_{xx}\Bigl(\hat{\sigma}_+\hat{\rho}_S\hat{\sigma}_--\hat{\rho}_S\hat{\sigma}_-\hat{\sigma}_+\Bigl)+ D_{pp}\Bigl(\hat{\sigma}_+\hat{\rho}_S\hat{\sigma}_-\nonumber\\
&-\hat{\sigma}_-\hat{\sigma}_+\hat{\rho}_S\Bigl)+ C_{pp}\Bigl(\hat{\sigma}_-\hat{\rho}_S\hat{\sigma}_+-\hat{\rho}_S\hat{\sigma}_+\hat{\sigma}_-\Bigl),
    \end{align}
\noindent
where $\hat{\sigma}_{\pm}$ are the Pauli raising and lowering operators for the qubit, satisfying the standard commutation relation $[\hat{\sigma}_{+},\hat{\sigma}_{-}]=\hat{\sigma}_{z}$.
%Using the same assumptions as in the quantum harmonic oscillator, we can rewrite the coefficients~(\ref{2lvmast}) as
Following the same procedure as before, the coefficients in Eq.~(\ref{2lvmast}) can be written as
\begin{align}\label{coeff2}
 D_{xx} &= \frac{1}{\hbar^2}\int_0^\infty d\tau \sum_n |C_n|^2\bigl\langle \hat{x}_n \hat{x}_n(-\tau) \bigl\rangle_B e^{i\Omega\tau},\nonumber\\
C_{xx} &= \frac{1}{\hbar^2}\int_0^\infty d\tau \sum_n |C_n|^2\bigl\langle \hat{x}_n(-\tau) \hat{x}_n \bigl\rangle_B e^{i\Omega\tau},\nonumber \\
D_{pp} &= \frac{1}{\hbar^2}\int_0^\infty d\tau\sum_n |C_n|^2\bigl\langle \hat{x}_n \hat{x}_n(-\tau) \bigl\rangle_B e^{-i\Omega\tau}, \nonumber \\
C_{pp} &=\frac{1}{\hbar^2} \int_0^\infty d\tau \sum_n |C_n|^2\bigl\langle \hat{x}_n(-\tau) \hat{x}_n \bigl\rangle_B e^{-i\Omega\tau}.
\end{align}
\noindent
%To evaluate Eq.~(\ref{coeff2}), we apply the rotating wave approximation (RWA)~\cite{thimmel1999rotating}, which amounts to disregarding the rapidly oscillating terms, and evaluate (assuming that $g_n=a_0C_n$, where $a_0$ denotes a relative coupling strength between the bath and the 2LS) the coefficients~(\ref{coeff2}) analytically to obtain
The coefficients in Eq.~(\ref{coeff2}) are obtained by evaluating the integrals, assuming $C_n = a_0 g_n$, where $a_0$ denotes the relative coupling strength between the bath and the 2LS, yielding
\begin{align}\label{coefn}
 D_{xx} &= C_{pp}^*=\alpha_1-i\alpha_-+i\delta_+,\nonumber\\
  D_{pp}&=C_{xx}^*=\alpha_2 -i\alpha_++i\delta_-,
 \end{align}
 where,
\begin{align}\label{2ls_params_1}
   &  \alpha_1 = \frac{2\pi}{\hbar} a_0^2J(\Omega)\bigl[n(\Omega)+1\bigl],\hspace{1mm} \delta_\pm = \frac{a_0^2}{\hbar}\mathrm{P}\int d\omega\,\frac{J(\omega)n(\omega)}{\omega\pm\Omega},\nonumber\\
   & \alpha_\pm = \frac{a_0^2}{\hbar}\mathrm{P}\int d\omega\,\frac{J(\omega)\bigl[n(\omega)+1\bigl]}{\omega\pm\Omega},\hspace{1mm}\alpha_2 = \frac{2\pi}{\hbar} a_0^2J(\Omega)n(\Omega).
 \end{align}
\noindent
Here, $\mathrm{P}$ denotes the Cauchy principal value. 
%In the high-temperature limit and large-cutoff limit $k_B T \gg \Lambda \gg \omega, \Omega$, Eq.~(\ref{2ls_params_1}) reduces to
%\begin{align}
 %   \alpha_1 &=\alpha_2\approx \frac{4m\gamma a_0^2 k_B T}{\hbar^2},\nonumber\\
%    \alpha_\pm &\approx \\ \nonumber
 %   \delta_\pm &\approx
%\end{align}
%\noindent
Inserting the coefficients~(\ref{coefn}) into the superoperator~(\ref{2lvmast}) leads to
\begin{align}\label{2lvsf00}
\pazocal{L}_{\mathrm{2LS}}\hat{\rho}_ S &= -\frac{i\Omega}{2}[\hat{\sigma}_z,\hat{\rho}_S] +2\alpha_1\mathcal{L}[\hat{\sigma}_-,\hat{\sigma}_+]\hat{\rho}_S\nonumber\\
     &+2\alpha_2\mathcal{L}[\hat{\sigma}_+,\hat{\sigma}_-]\hat{\rho}_S+i[\alpha_-\hat{\sigma}_+\hat{\sigma}_--\delta_-\hat{\sigma}_-\hat{\sigma}_+,\hat{\rho}_S]\nonumber\\
     &-i[\delta_+\hat{\sigma}_+\hat{\sigma}_--\alpha_+\hat{\sigma}_-\hat{\sigma}_+,\hat{\rho}_S],
\end{align}
\noindent
where the superoperator $\pazocal{L}[\hat{y},\hat{y}^\dagger]\hat{\rho}_S=\hat{y}\hat{\rho}_S \hat{y}^\dagger-(1/2)\{\hat{y}^\dagger\hat{y},\hat{\rho}_S\}$ denotes the standard Gorini--Kossakowski--Sudarshan--Lindblad (GKSL) dissipator~\cite{lindblad1976generators,gorini1976completely}. 
Equation~(\ref{2lvsf00}) can be written in the simple form as 
\begin{align}\label{2lvsf}
\pazocal{L}_{\mathrm{2LS}}\hat{\rho}_ S =\ &i\bar{\lambda}_1[\hat{\sigma}_z,\hat{\rho}_S] +\bar{\lambda}_2\mathcal{L}[\hat{\sigma}_-,\hat{\sigma}_+]\hat{\rho}_S\nonumber\\
     &+\bar{\lambda}_3\mathcal{L}[\hat{\sigma}_+,\hat{\sigma}_-]\hat{\rho}_S,
\end{align}
where,
\begin{align}\label{tls_params}
\bar{\lambda}_1 &= \frac{a_0^2}{2\hbar}\mathrm{P}\int d\omega\, J(\omega)\bigl[2n(\omega)+1\bigl]\biggl(\frac{ 1}{\omega-\Omega}-\frac{1}{\omega+\Omega}\biggl)-\frac{\Omega}{2},\nonumber\\
    \bar{\lambda}_2 &= 2\Gamma(\Omega)\bigl[n(\Omega)+1\bigl],\hspace{4mm}
     \bar{\lambda}_3 = 2\Gamma(\Omega)n(\Omega).
\end{align}
\noindent
Here, $n(\Omega)$ denotes the Planck distribution at the transition frequency $\Omega$ and $\Gamma(\Omega)=2a_0^2\pi J(\Omega)/\hbar$ is the spontaneous emission rate. %Equation~(\ref{2lvsf}) is a well-known quantum optical master equation for the 2LS~\cite{carmichael1999statistical,scully1999quantum}.
Equation~(\ref{2lvsf}) is a well-known quantum optical superoperator governing the dynamics of the 2LS~\cite{carmichael1999statistical,scully1999quantum}. In the high-temperature limit and large-cutoff limit $k_B T \gg \hbar\Lambda \gg \hbar\Omega$, the coefficients~(\ref{tls_params}) reduce to
\begin{align}
    \bar{\lambda}_1 \approx -\biggl(\frac{2\Omega}{\Lambda}\biggl)\frac{m\gamma}{\hbar^2} k_B Ta_0^2, \hspace{4mm}\bar{\lambda}_{2,3} \approx 8\frac{m\gamma}{\hbar^2}k_B Ta_0^2.
\end{align}
\noindent
Since $\bar{\lambda}_2 \approx \bar{\lambda}_3$ in the high-temperature and large-cutoff limits, we henceforth use $\bar{\lambda}_2$ to denote both rates.
Next, we consider the ``cross-term''~(\ref{brw2}). By using Eq.~(\ref{transtau}) together with
\begin{align}
    \hat{x}(-\tau) &=x_0\bigl(\hat{a}e^{i\omega\tau} +  \hat{a}^\dagger e^{-i\omega\tau}\bigl),\nonumber\\
    \hat{x} &= x_0\bigl(\hat{a}+\hat{a}^\dagger\bigl),
\end{align}
\noindent
and proceeding as previously, one can show that Eq.~(\ref{brw2}) can be written as
\begin{align}\label{croscoff}
&\pazocal{L}_{\mathrm{cross}}\hat{\rho}_S = D_{xy}\Bigl(\hat{\sigma}_-\hat{\rho}_S\hat{a}^\dagger-\hat{a}^\dagger\hat{\sigma}_-\hat{\rho}_S\Bigl)\nonumber\\
&+C_{xy}\Bigl(\hat{\sigma}_+\hat{\rho}_S\hat{a}-\hat{a}\hat{\sigma}_+\hat{\rho}_S\Bigl)+D_{zy}\Bigl(\hat{a}\hat{\rho}_S\hat{\sigma}_+-\hat{\sigma}_+\hat{a}\hat{\rho}_S\Bigl)\nonumber\\
&+C_{zy}\Bigl(\hat{a}^\dagger\hat{\rho}_S\hat{\sigma}_--\hat{\sigma}_-\hat{a}^\dagger\hat{\rho}_S\Bigl)
   + \mathrm{h.c}.
\end{align}   
\noindent
Here, $\mathrm{h.c.}$ denotes the Hermitian conjugate, $\hat{a}^{\dag}$ and $\hat{a}$  are the bosonic creation and annihilation operators for the harmonic-oscillator quanta, satisfying the fundamental bosonic commutation relations $[\hat{a}_k,\hat{a}_l^\dagger]=\delta_{k,l}$. The coefficients appearing in Eq.~(\ref{croscoff}) are 
\begin{align}\label{coefnc}
D_{xy} &= \frac{1}{\hbar^2}\int_0^\infty d\tau \sum_n g_n C_nx_0\bigl\langle \hat{x}_n \hat{x}_n(-\tau) \bigl\rangle_B e^{i\Omega\tau},\nonumber\\
C_{xy} &=\frac{1}{\hbar^2} \int_0^\infty d\tau \sum_n g_n C_nx_0\bigl\langle \hat{x}_n \hat{x}_n(-\tau) \bigl\rangle_B e^{-i\Omega\tau},\nonumber\\
D_{zy} &=\frac{1}{\hbar^2} \int_0^\infty d\tau \sum_n g_n C_nx_0\bigl\langle \hat{x}_n \hat{x}_n(-\tau) \bigl\rangle_B e^{i\omega\tau},\nonumber\\
C_{zy} &=\frac{1}{\hbar^2} \int_0^\infty d\tau \sum_n g_n C_nx_0\bigl\langle \hat{x}_n \hat{x}_n(-\tau) \bigl\rangle_B e^{-i\omega\tau}.
\end{align} 
\noindent
To evaluate these coefficients~(\ref{coefnc}), we assume that the quantum oscillator and the 2LS are at resonance, i.e., $\omega=\Omega$ ($g_n=a_0C_n$). %Again, we apply the RWA, which leads to
As before, analytic integration yields
\begin{align}\label{coeffcrosst}
 D_{xy} &=D_{zy} =\beta_1-i\beta_-+i\mu_+,\nonumber\\
 C_{xy} &=C_{zy}=\beta_2-i\beta_++i\mu_-,
 \end{align}
 where,
 \begin{align}
   &  \beta_1 = \frac{2\pi}{\hbar} a_0x_0J(\Omega)\bigl[n(\Omega)+1\bigl],\hspace{2mm} \beta_2 = \frac{2\pi}{\hbar} a_0x_0J(\Omega)n(\Omega),\nonumber\\
   & \beta_\pm = \frac{a_0x_0}{\hbar}\mathrm{P}\int d\omega'\,\frac{J(\omega')\bigl[n(\omega')+1\bigl]}{\omega'\pm\Omega},\nonumber\\
   & \mu_\pm = \frac{a_0x_0}{\hbar}\mathrm{P}\int d\omega'\,\frac{J(\omega')n(\omega')}{\omega'\pm\Omega}.
 \end{align}
\noindent
%After some algebra, one can now show that Eq.~(\ref{brw2}) reduces to
Inserting the coefficients~(\ref{coeffcrosst}) into the superoperator~(\ref{croscoff}) leads to
\begin{align}\label{eqfin1}
  & \pazocal{L}_{\mathrm{cross}}\hat{\rho}_S = \bar{\beta}_1\Bigl(\mathcal{L}[\hat{a},\hat{\sigma}_+]\hat{\rho}_S+\mathcal{L}[\hat{\sigma}_-,\hat{a}^\dagger]\hat{\rho}_S\nonumber+\mathcal{L}[\hat{a}^\dagger,\hat{\sigma}_-]\hat{\rho}_S\nonumber\\
  &+\mathcal{L}[\hat{\sigma}_+,\hat{a}]\hat{\rho}_S\Bigl)+\bar{\beta}_2\Bigl(\mathcal{L}[\hat{a},\hat{\sigma}_+]\hat{\rho}_S+\mathcal{L}[\hat{\sigma}_-,\hat{a}^\dagger]\hat{\rho}_S\Bigl)\nonumber\\
  &+i\bar{\beta}_3\bigl[\hat{a}^\dagger\hat{\sigma}_-+\hat{a}\hat{\sigma}_+,\hat{\rho}_S\bigl],
\end{align}
where,
\begin{align}\label{cross_params}
   &  \bar{\beta}_1 = \frac{4\pi}{\hbar} a_0x_0J(\Omega)n(\Omega),\hspace{2mm} \bar{\beta}_2 = \frac{4\pi}{\hbar} a_0x_0J(\Omega),\nonumber\\
   & \bar{\beta}_3 = \frac{a_0x_0}{\hbar}\mathrm{P}\int d\omega'\,J(\omega')\biggl(\frac{1}{\omega'-\Omega}+\frac{1}{\omega'+\Omega}\biggl).
 \end{align}
\noindent
In the high-temperature limit and large-cutoff limit, where $k_B T \gg \hbar\Lambda \gg \hbar\Omega$, the coefficients~(\ref{cross_params}) simplifly to
\begin{align}
    \bar{\beta}_1 &\approx \frac{8m\gamma}{\hbar^2} k_B Ta_0x_0,\hspace{4mm} \bar{\beta}_2 \approx \frac{8m\gamma}{\hbar}\Omega a_0x_0,\nonumber\\
    \bar{\beta}_3 &\approx \frac{2m\gamma}{\hbar} \Lambda a_0x_0.
\end{align}
\noindent
Combining the dissipators of the quantum harmonic oscillator~(\ref{qhomastfnf}), the 2LS (\ref{2lvsf}) and the ``cross-term'' dissipator~(\ref{eqfin1}), and substituting $x_0$ and $p_0$, we end up with
\begin{align}\label{fin_meq}
&\frac{d}{d t} \hat{\rho}_S (t) = -\frac{i\omega}{4}\bigl[\hat{p}^2+\hat{x}^2,\hat{\rho}_S\bigl] -\gamma\biggl(\frac{k_B T}{\hbar \omega}\biggl)\bigl[\hat{x},[\hat{x},\hat{\rho}_S]\bigl] \nonumber\\
&-\frac{i\gamma}{2} \bigl[\hat{x},\{\hat{p},\hat{\rho}_S\}\bigl]+ i\bar{\lambda}_1[\hat{\sigma}_z,\hat{\rho}_S] +\bar{\lambda}_2\mathcal{L}[\hat{\sigma}_-,\hat{\sigma}_+]\hat{\rho}_S\nonumber
\\
&+\bar{\lambda}_2\mathcal{L}[\hat{\sigma}_+,\hat{\sigma}_-]\hat{\rho}_S+\bar{\beta}_1\Bigl(\mathcal{L}[\hat{a},\hat{\sigma}_+]\hat{\rho}_S+\mathcal{L}[\hat{\sigma}_-,\hat{a}^\dagger]\hat{\rho}_S\nonumber\\
&+\mathcal{L}[\hat{a}^\dagger,\hat{\sigma}_-]\hat{\rho}_S+\mathcal{L}[\hat{\sigma}_+,\hat{a}]\hat{\rho}_S\Bigl)+\bar{\beta}_2\Bigl(\mathcal{L}[\hat{a},\hat{\sigma}_+]\hat{\rho}_S\nonumber\\
&+\mathcal{L}[\hat{\sigma}_-,\hat{a}^\dagger]\hat{\rho}_S\Bigl)+i\bar{\beta}_3\bigl[\hat{a}^\dagger\hat{\sigma}_-+\hat{a}\hat{\sigma}_+,\hat{\rho}_S\bigl].
\end{align}
\noindent
In the next section, we demonstrate a systematic method to eliminate the momentum variable from Eq.~(\ref{fin_meq}) via adiabatic elimination and derive the master equation for the OQBM.

%%%%%%%%%%%%%%%%%%%%%%%%%%%%%%%%%%%%%%%%%%%%%%%%%%%%%%
\section{Adiabatic elimination of the momentum variable}\label{part_2}
%%%%%%%%%%%%%%%%%%%%%%%%%%%%%%%%%%%%%%%%%%%%%%

Adiabatic elimination is a standard technique used to derive an effective dynamics for slow variables by eliminating fast-evolving ones~\cite{SKINNER1979561,van1985elimination,kramers1940brownian,gardiner1985handbook}.
In the model considered in this paper, the momentum of the Brownian particle under the assumption of the large damping dissipates very fast, 
whereas its position changes considerably more gradually. In other words, the momentum and the position evolve on different time scales. The dynamics of the Brownian particle may then be described using position variables alone by adiabatically eliminating the rapidly relaxing momentum. 

For the problem under consideration, we assume that the damping coefficient $\gamma$ is very large compared to all the parameters of the system, i.e., $\omega,\Omega \ll \gamma$.
The damping coefficient represents the timescale for a fast momentum variable. The projection operator method systematically splits the momentum and position dynamics into their fast and slow components. 
To obtain the behavior of interest, we use the method developed in~\cite{gardiner1985handbook} to perform adiabatic elimination of the momentum variable from Eq.~(\ref{fin_meq}).

To demonstrate this procedure, we transform Eq.~(\ref{fin_meq}) into the phase-space representation using the Wigner function~\cite{wigner1932quantum,hillery1984distribution} and then perform the adiabatic elimination of the momentum variable. 
The master equation in the phase-space representation for the position variable describes the OQBM.

%%%%%%%%%%%%%%%%%%%%%%%%%%%%%%%%%%%%%%%%%%%%%%%%%%%%
\subsection{Master equation for the Wigner function}
%%%%%%%%%%%%%%%%%%%%%%%%%%%%%%%%%%%%%%%%%%%%%%%%%%%%
The quantum master equation for the reduced density matrix $\hat{\rho}_S (t)$ can be written in 
terms of the Wigner function $\hat{W}(x,p,t)$. The Wigner function represents a quasi-probability 
distribution of the density matrix in phase-space. Equation~(\ref{fin_meq}) can be transformed by using the following relations~\cite{schleich2001,gardiner2004quantum}:
\begin{align}\label{wignertr}
    \hat{x}\hat{\rho} & \leftrightarrow \biggl(x+i\frac{\partial}{\partial p} \biggl)\hat{W}, \hspace{4mm} \hat{\rho} \hat{x} \leftrightarrow \biggl(x-i\frac{\partial}{\partial p} \biggl)\hat{W}, \nonumber\\
     \hat{p}\hat{\rho} &\leftrightarrow \biggl(p-i\frac{\partial}{\partial x} \biggl)\hat{W}, \hspace{4mm} \hat{\rho}\hat{p} \leftrightarrow \biggl(p+i\frac{\partial}{\partial x} \biggl)\hat{W}.
\end{align}
\noindent
By using Eq.~(\ref{wignertr}) and 
\begin{align}
\hat{W}(x,p,t) &= \frac{1}{4\pi}\int dy\, e^{-ipy/2} \langle x+  y/2|\hat{\rho}| x- y/2\rangle,\nonumber\\
\langle x | p \rangle &= \frac{1}{\sqrt{4\pi }} e^{i px /2},
\end{align}
\noindent
one can show that Eq.~(\ref{fin_meq}) in phase-space simplifies to 
\begin{align}\label{weqfin1}
\pazocal{L}_{\mathrm{QHO}}\hat{W} =\ & 4\gamma\biggl(\frac{k_B T}{\hbar \omega}\biggl)\frac{\partial^2}{\partial p^2}\hat{W}+2\gamma\frac{\partial}{\partial p} \bigl(p\hat{W}\bigl)+x\omega\frac{\partial}{\partial p} \hat{W}\nonumber\\
&-p\omega\frac{\partial}{\partial x}\hat{W},\nonumber\\
 \pazocal{L}_{\mathrm{2LS}}\hat{W}=\ &i\bar{\lambda}_1[\hat{\sigma}_z,\hat{W}]+ \bar{\lambda}_2\mathcal{L}[\hat{\sigma}_-,\hat{\sigma}_+]\hat{W}\nonumber\\
 &+\bar{\lambda}_2\mathcal{L}[\hat{\sigma}_+,\hat{\sigma}_-]\hat{W},\nonumber\\
\pazocal{L}_{\mathrm{cross}}\hat{W}  = \ &  \biggl(\frac{\partial}{\partial p} \hat{\mathcal{M}}_1+\frac{\partial}{\partial x} \hat{\mathcal{M}}_2+ x \hat{\mathcal{M}}_3+p \hat{\mathcal{M}}_4\biggl)\hat{W}.
\end{align}
\noindent
%Here, $\gamma$ is a large parameter, and we employ $1/\gamma$ as a small parameter to eliminate the fast variable $p$. The  super-operators $\hat{m}_1$, $\hat{m}_2$, $\hat{m}_3$, and $\hat{m}_4$ acting on the internal degree of freedom must not be confused with the ``mass'' and are respectively given by
Here, $\gamma$ is a large parameter, and we employ $1/\gamma$ as a small parameter to eliminate the fast variable $p$. The superoperators $\hat{\mathcal{M}}_1$, $\hat{\mathcal{M}}_2$, $\hat{\mathcal{M}}_3$, and $\hat{\mathcal{M}}_4$ acting on the internal degree of freedom are respectively given by
\begin{align}
    \hat{\mathcal{M}}_1 &= -\frac{\bar{\beta}_2}{4}\Bigl(\{\hat{\sigma}_y,\cdot\}+2i[\hat{\sigma}_x,\cdot\ ]\Bigl)-\frac{\bar{\beta}_3}{2}\{\hat{\sigma}_x,\cdot \}-i\bar{\beta}_1[\hat{\sigma}_x,\cdot\ ],\nonumber\\
      \hat{\mathcal{M}}_2 &= \frac{\bar{\beta}_2}{4}\Bigl(\{\hat{\sigma}_x,\cdot \}-2i[\hat{\sigma}_y,\cdot \ ]\Bigl)-\frac{\bar{\beta}_3}{2}\{\hat{\sigma}_y,\cdot \}-i\bar{\beta}_1[\hat{\sigma}_y,\cdot \ ] ,\nonumber\\
        \hat{\mathcal{M}}_3 &= -i\frac{\bar{\beta}_2}{4}[\hat{\sigma}_y,\cdot \ ]+i\frac{\bar{\beta}_3}{2}[\hat{\sigma}_x,\cdot \ ],\nonumber\\
          \hat{\mathcal{M}}_4 &=-i\frac{\bar{\beta}_2}{4}[\hat{\sigma}_x,\cdot \ ]-i\frac{\bar{\beta}_3}{2}[\hat{\sigma}_y,\cdot \ ].
\end{align}
\noindent
We can combine Eq.~(\ref{weqfin1}) as in Eq.~(\ref{mastersep}) and write it in the following compact form:
\begin{align}\label{weqfin2}
   \frac{\partial }{\partial t} \hat{W}&=  \pazocal{L}_{\mathrm{QHO}}\hat{W}+\biggl(\frac{\partial}{\partial p} \hat{\mathcal{M}}_1+\frac{\partial}{\partial x} \hat{\mathcal{M}}_2+ x \hat{\mathcal{M}}_3\nonumber\\
   &\quad+p \hat{\mathcal{M}}_4\biggl)\hat{W} +\pazocal{L}_{\mathrm{2LS}} \hat{W}.
\end{align}
\noindent
From Eq.~(\ref{weqfin1}), it is clear that the following commutator holds
$\bigl[\pazocal{L}_{\mathrm{QHO}}, \pazocal{L}_{\mathrm{2LS}}\bigl] =0.
$
We now proceed to eliminate the fast-relaxing momentum variable adiabatically. 
This technique has already been established and used in the derivation of 
the famous Smoluchowski equation~\cite{smoluchowski1916brownsche}, see Ref.~\cite{gardiner1985handbook} Section 6.4 for 
a more detailed discussion.
Our main goal is to write Eq.~(\ref{weqfin2}) as a function of the 
position variable.  

%%%%%%%%%%%%%%%%%%%%%%%%%%%%%%%%%%%%%%%%%%%%%%%%%%%%%%%%%%%%%
\subsection{General formulation in terms of operators and projectors}
%%%%%%%%%%%%%%%%%%%%%%%%%%%%%%%%%%%%%%%%%%%%%%%%%%%%%%%%%%%%%
In the following, we describe our elimination method.
As a first step, we write Eq.~(\ref{weqfin2}) as
\begin{align}\label{weqfin3}
   \frac{\partial}{\partial t} \hat{W} &= \bigl(\gamma \hat{L}_1 +\hat{L}_2\bigl)\hat{W}+\biggl(\frac{\partial}{\partial p} \hat{\mathcal{M}}_1+\frac{\partial}{\partial x} \hat{\mathcal{M}}_2+ x \hat{\mathcal{M}}_3\nonumber\\
   &\quad+p \hat{\mathcal{M}}_4\biggl)\hat{W} +\pazocal{L}_{\mathrm{2LS}} \hat{W},
\end{align}
\noindent
where, $\hat{L}_1$ and $\hat{L}_2$ are 
  \begin{align}
     \hat{L}_1 & = \alpha \frac{\partial^2}{\partial p^2} + 2\frac{\partial}{\partial p} p, \label{eq_1}\\
    \hat{L}_2 & = -p\omega\frac{\partial}{\partial x}+u(x) \frac{\partial}{\partial p}. \label{eq_2}
\end{align}
\noindent
Here, $\alpha=4k_B T/ \hbar\omega$ and $u(x)= x\omega $. The operator $\hat{L}_1$ describes the momentum distribution's relaxation on the time scale $\gamma^{-1}$. We are looking for the position distribution function for $x$, $\bar{W}(x, t)$, defined by
\begin{equation}
\bar{W}(x, t) = \int_{-\infty}^{+\infty} dp\, \hat{W}(x,p,t).
\end{equation}
\noindent
For large $\gamma$, the momentum distribution is rapidly thermalized, and the spatial distribution obeys a diffusion equation.
The stationary distribution of Eq.~(\ref{weqfin3}) can be obtained by solving the differential operator~(\ref{eq_1}), which reads
\begin{equation}\label{a111}
     \frac{\partial}{\partial t} \hat{W} = \hat{L}_1\hat{W}=\alpha \frac{\partial^2 }{\partial p^2}\hat{W} +  2\frac{\partial }{\partial p}(p\hat{W})=0.
\end{equation}
\noindent
It is straightforward to show that the solution of Eq.~(\ref{a111}) is 
\begin{equation}
    w_s(p)=(\pi\alpha)^{-1/2}\exp(-p^2/\alpha).
\end{equation}
\noindent
In the next step, we introduce a projection operator $\mathcal{P}$, defined as
\begin{equation}\label{aa1}
\mathcal{P}f(p,x) = w_s(p)\int dp'\,f(p',x),
\end{equation}
\noindent
where $f(p,x)$ is an arbitrary function. This operator $\mathcal{P}$ satisfies $\mathcal{P}^2=\mathcal{P}$, and it works as the projection operator to the relevant part of the total Wigner function $\hat{W}(x,p,t)$. 
The results of applying $\mathcal{P}$ to $\hat{W}(x,p,t)$ yields
\begin{equation}\label{pw}
\mathcal{P}\hat{W}(x,p,t) = w_s(p)\bar{W}(x,t).
\end{equation}
\noindent
Formally, Eq.~(\ref{aa1}) can be written as
\begin{equation}\label{aa12a}
g(p,x) = w_s(p)\hat{g}(x),
\end{equation}
\noindent
where $g(p,x)$ is an arbitrary function.
On the other hand, functions of type~(\ref{aa12a}) are all solutions of
\begin{equation}
   \hat{L}_1 g = 0,
\end{equation}
\noindent
that is, the space that $\mathcal{P}$ projects onto is the null space of $\hat{L}_1$. Consequently, the projector $\mathcal{P}$ can be written as
\begin{equation}\label{eqw}
\mathcal{P} = \lim_{t\to\infty}\bigl[\exp\bigl(\hat{L}_1 t\bigl)\bigl].
\end{equation}
\noindent
To verify Eq.~(\ref{eqw}), we can expand any function of $p$ and $x$ in eigenfunctions $P_\lambda(p)$ of $\hat{L}_1$ (see Eq.~(\ref{bookeqg})) as
\begin{align}
f(p,x) &= \sum_{\lambda} A_\lambda (x) P_\lambda(p), \\
\text{where,} \hspace{4mm} A_\lambda(x) &= \int dp\, Q_\lambda (p) f(p,x).
\end{align}
\noindent
Here, $Q_\lambda (p)$ are the dual eigenfunctions corresponding to the eigenfunctions $P_\lambda(p)$ of the operator $\hat{L}_1$, forming a biorthogonal set used to project onto the eigenmodes of the fast momentum variable.
Then, the long-time limit can be expressed as
\begin{align}
\lim_{t\to\infty}\bigl[\exp\bigl(\hat{L}_1 t\bigl) f(p,x)\bigl] & =\sum_{\lambda} A_\lambda (x)  \lim_{t\to\infty}e^{-\lambda t} P_\lambda(p)\nonumber\\
 &=P_0(p)\int dp\, Q_0(p) f(p,x),
\end{align}
\noindent
where,
\begin{align}
P_0(p) =(\pi\alpha)^{-1/2}\exp(-p^2/\alpha),\hspace{4mm}
 Q_0(p)=1. 
\end{align}
\noindent
In this case and all other cases, we also have the following crucial relation $\mathcal{P}\hat{L}_2 \mathcal{P} = 0$,
and noting that for this process
\begin{equation}
p\exp(-p^2/\alpha) \propto P_1(p),   \hspace{3mm} \text{and} \hspace{3mm} \mathcal{P}P_1(p)=0.
\end{equation}
\noindent
We define $\mathcal{Q}=1-\mathcal{P}$, where the operators $\mathcal{P}$ and $\mathcal{Q}$ select the relevant and the irrelevant part of $\hat{W}(x,p,t)$, respectively. The standard properties of projectors $\mathcal{Q}^2=\mathcal{Q}$ and $\mathcal{P}\mathcal{Q}=\mathcal{Q}\mathcal{P}=0$, hold. Following the projection operator formalism, we can write
\begin{align}
v &= \mathcal{P}\hat{W}, \\
w&=(1-\mathcal{P})\hat{W}.
\end{align}
\noindent
Here, the function $v= P_0(p) \bar{W}$ plays the role of slow variables, and $w$ plays the role of fast variables. Consequently, $\hat{W}$ can now be decomposed into two parts
\begin{equation}\label{weqfin4}
    \hat{W} = v+w.
\end{equation}
\noindent
Also, from Eq.~(\ref{eqw}) it is clear that $\mathcal{P}\hat{L}_1=\hat{L}_1\mathcal{P}=0$. Applying the projection operators $\mathcal{P}$ and $(1-\mathcal{P})$ to Eq.~(\ref{weqfin3}), we obtain
\begin{align}
   \frac{\partial v}{\partial t} = \ &  \mathcal{P}\bigl(\gamma \hat{L}_1 +\hat{L}_2\bigl)\hat{W}+\mathcal{P}\biggl(\frac{\partial}{\partial p} \hat{\mathcal{M}}_1+\frac{\partial}{\partial x} \hat{\mathcal{M}}_2+ x \hat{\mathcal{M}}_3\nonumber\\
   &+p \hat{\mathcal{M}}_4\biggl)\hat{W} +\mathcal{P}\pazocal{L}_{\mathrm{2LS}} \hat{W},\nonumber\\
   \frac{\partial w}{\partial t} =\ &  \mathcal{Q}\bigl(\gamma \hat{L}_1 +\hat{L}_2\bigl)\hat{W}+\mathcal{Q}\biggl(\frac{\partial}{\partial p} \hat{\mathcal{M}}_1+\frac{\partial}{\partial x} \hat{\mathcal{M}}_2+ x \hat{\mathcal{M}}_3\nonumber\\
   &+p \hat{\mathcal{M}}_4\biggl)\hat{W}+\mathcal{Q}\pazocal{L}_{\mathrm{2LS}} \hat{W}.
\end{align}
\noindent 
After some algebra, and by using $\mathcal{P}\hat{L}_2\mathcal{P}=0$, 
which can be verified by a straightforward brute-force calculation, along with 
$\mathcal{P} \pazocal{L}_{\mathrm{2LS}}\mathcal{P}=\pazocal{L}_{\mathrm{2LS}}\mathcal{P}=\mathcal{P} \pazocal{L}_{\mathrm{2LS}}$, we obtain
\begin{align}\label{abc11}
 \frac{\partial v}{\partial t}&= \mathcal{P}\hat{L}_2w + \hat{\mathcal{M}}_2 \frac{\partial v}{\partial x} +x\hat{\mathcal{M}}_3 v +\mathcal{P}\bigl(p\hat{\mathcal{M}}_4w\bigl)+\pazocal{L}_{\mathrm{2LS}}v,\nonumber\\
 \frac{\partial w}{\partial t} &= \gamma \hat{L}_1w+(1-\mathcal{P})\hat{L}_2w +\hat{L}_2v+\hat{\mathcal{M}}_1\frac{\partial v}{\partial p} + \hat{\mathcal{M}}_1\frac{\partial w} {\partial p}\nonumber\\
 &\quad+ \hat{\mathcal{M}}_2\frac{\partial w}{\partial x}+x\hat{\mathcal{M}}_3w+p\hat{\mathcal{M}}_4v+p\hat{\mathcal{M}}_4w\nonumber\\
 &\quad-\mathcal{P}\bigl(p\hat{\mathcal{M}}_4w\bigl)+\pazocal{L}_{\mathrm{2LS}}w.
\end{align}

%\vspace{2mm}
%%%%%%%%%%%%%%%%%%%%%%%%%%%%%%%%%%%%%%%%%%%%%%%%%%%%%%%%%%%%%
\subsection{Solution using Laplace transform}
%%%%%%%%%%%%%%%%%%%%%%%%%%%%%%%%%%%%%%%%%%%%%%%%%%%%%%%%%%%%%
Here, we solve Eq.~(\ref{abc11}) using the Laplace transform:
\begin{equation}\label{laplace}
\tilde{f}(s) = \int_0^\infty dt\, e^{-st} f(t),  
\end{equation}
\noindent
where $f(t)$ is an arbitrary function of time. This transformation~(\ref{laplace}) applied to Eq.~(\ref{abc11}) yields
\begin{align}
s\tilde{v}(s)  &=  \mathcal{P}\hat{L}_2\tilde{w}(s)+ \hat{\mathcal{M}}_2 \frac{\partial \tilde{v}(s)}{\partial x} +x\hat{\mathcal{M}}_3 \tilde{v}(s) +\pazocal{L}_{\mathrm{2LS}}\tilde{v}(s)\nonumber\\
&+\mathcal{P}\bigl(p\hat{\mathcal{M}}_4\tilde{w}(s)\bigl)+v(0), \label{cn1}\\
s\tilde{w}(s) &= \gamma \hat{L}_1\tilde{w}(s)+(1-\mathcal{P})\hat{L}_2\tilde{w}(s) +\hat{L}_2\tilde{v}(s)+\hat{\mathcal{M}}_1\frac{\partial \tilde{v}(s)}{\partial p} \nonumber\\
&+ \hat{\mathcal{M}}_1\frac{\partial \tilde{w}(s)}{\partial p}+ \hat{\mathcal{M}}_2\frac{\partial \tilde{w}(s)}{\partial x}+x\hat{\mathcal{M}}_3\tilde{w}(s)+p\hat{\mathcal{M}}_4\tilde{v}(s)\nonumber\\
&+p\hat{\mathcal{M}}_4\tilde{w}(s)-\mathcal{P}\bigl(p\hat{\mathcal{M}}_4\tilde{w}(s)\bigl)+\pazocal{L}_{\mathrm{2LS}}\tilde{w}(s)+w(0).\label{abcs}
\end{align}
\noindent
We assume that $w(0)=0$, which means that the initial distribution is assumed to be of the form 
\begin{equation}
    \hat{W}(x,p,0) = (\pi\alpha)^{-1/2}\exp(-p^2/\alpha)\bar{W}(x,0),
\end{equation}
\noindent 
which allows us to satisfy the condition of initial thermalization of the momentum variable. We solve Eq.~(\ref{abcs}) for $\tilde{w}(s)$ to obtain
\begin{align}\label{abcs1}
\tilde{w}(s)&=\bigl[s-\gamma \hat{L}_1 -(1-\mathcal{P})\hat{L}_2-\hat{\mathcal{M}}_1\frac{\partial}{\partial p}-\hat{\mathcal{M}}_2\frac{\partial}{\partial x}-x\hat{\mathcal{M}}_3\nonumber\\
&\quad-p\hat{\mathcal{M}}_4+\mathcal{P}\bigl(p\hat{\mathcal{M}}_4\bigl)-\pazocal{L}_{\mathrm{2LS}}\bigl]^{-1}\nonumber\\
&\quad
\times \biggl(\hat{L}_2+p\hat{\mathcal{M}}_4+\hat{\mathcal{M}}_1\frac{\partial}{\partial p}\biggl)\tilde{v}(s).
\end{align}
\noindent
We substitute Eq.~(\ref{abcs1}) into Eq.~(\ref{cn1}) to find
\begin{widetext}
\begin{align}\label{cn1q}
s\tilde{v}(s)-v(0)&=  \Bigl[\mathcal{P}\hat{L}_2+\mathcal{P}\bigl(p\hat{\mathcal{M}}_4\bigl)\Bigl]\Bigl[s-\gamma \hat{L}_1 -(1-\mathcal{P})\hat{L}_2-\hat{\mathcal{M}}_1\frac{\partial}{\partial p}-\hat{\mathcal{M}}_2\frac{\partial}{\partial x}-x\hat{\mathcal{M}}_3-p\hat{\mathcal{M}}_4+\mathcal{P}\bigl(p\hat{\mathcal{M}}_4\bigl)-\pazocal{L}_{\mathrm{2LS}}\Bigl]^{-1}\nonumber\\
&\quad\times \biggl(\hat{L}_2+p\hat{\mathcal{M}}_4+\hat{\mathcal{M}}_1\frac{\partial}{\partial p}\biggl)\tilde{v}(s)+\hat{\mathcal{M}}_2\frac{\partial \tilde{v}(s)}{\partial x}+x\hat{\mathcal{M}}_3\tilde{v}(s)+\pazocal{L}_{\mathrm{2LS}}\tilde{v}(s).
\end{align}
\end{widetext}

\noindent
Here, we have a partial solution to the problem. For any finite $s$,  we take the large $\gamma $ limit of Eq.~(\ref{cn1q}) to obtain
\begin{align}
&s\tilde{v}(s)-v(0) \approx -\gamma^{-1}\mathcal{P}\hat{L}_2 \hat{L}_1^{-1}\hat{L}_2\tilde{v}(s)+\hat{\mathcal{M}}_2\frac{\partial \tilde{v}(s)}{\partial x}\nonumber\\
&-\gamma^{-1}\mathcal{P}\hat{L}_2\hat{L}_1^{-1}\biggl(p\hat{\mathcal{M}}_4+\hat{\mathcal{M}}_1\frac{\partial}{\partial p}\biggl)\tilde{v}(s)+ x\hat{\mathcal{M}}_3\tilde{v}(s)\nonumber\\
&-\gamma^{-1}\mathcal{P}p\hat{\mathcal{M}}_4\hat{L}_1^{-1}\biggl(p\hat{\mathcal{M}}_4+\hat{\mathcal{M}}_1\frac{\partial}{\partial p}\biggl)\tilde{v}(s)+\pazocal{L}_{\mathrm{2LS}}\tilde{v}(s)\nonumber\\
&-\gamma^{-1}\mathcal{P}p\hat{\mathcal{M}}_4\hat{L}_1^{-1}\hat{L}_2\tilde{v}(s).
\end{align}
\noindent
Taking the inverse Laplace transform gives
\begin{align}\label{felimp}
 \frac{\partial v}{\partial t}= &-\gamma^{-1}\mathcal{P}\hat{L}_2 \hat{L}_1^{-1}\hat{L}_2 v+\hat{\mathcal{M}}_2\frac{\partial v}{\partial x}+x\hat{\mathcal{M}}_3v+\pazocal{L}_{\mathrm{2LS}}v \nonumber\\
    &-\gamma^{-1}\mathcal{P}\hat{L}_2 \hat{L}_1^{-1} \biggl(p\hat{\mathcal{M}}_4+\hat{\mathcal{M}}_1\frac{\partial}{\partial p}\biggl)v\nonumber\\
    &-\gamma^{-1}\mathcal{P}p\hat{\mathcal{M}}_4\hat{L}_1^{-1}\biggl(p\hat{\mathcal{M}}_4+\hat{\mathcal{M}}_1\frac{\partial}{\partial p}\biggl)v\nonumber\\
    &-\gamma^{-1}\mathcal{P}p\hat{\mathcal{M}}_4\hat{L}_1^{-1}\hat{L}_2v.
\end{align}
\noindent
The operators $\mathcal{P}\hat{L}_2 \hat{L}_1^{-1}\hat{L}_2 v$, $\mathcal{P}\hat{L}_2 \hat{L}_1^{-1}(\cdot)v$, $\mathcal{P}p\hat{\mathcal{M}}_4\hat{L}_1^{-1}(\cdot)$, and $\mathcal{P}p\hat{\mathcal{M}}_4\hat{L}_1^{-1}\hat{L}_2v$ are evaluated next (see Appendix~\ref{appendixa} for details).
%By direct substitution of $v=\mathcal{P}\hat{W}=P_0(p)\bar{W}$ into the master equation~(\ref{felimp}) and neglecting terms that scales as an order of $||\hat{m}_i||/\gamma$ we obtain the following equation
By directly substituting $v=\mathcal{P}\hat{W}=P_0(p)\bar{W}$ into the master equation~(\ref{felimp}), we obtain the leading-order $1/\gamma$ equation in the overdamped expansion: 
\begin{align}\label{finmee}
 \frac{\partial}{\partial t} \bar{W}(x,t) &= \bar{\alpha}\frac{\partial^2 }{\partial x^2}\bar{W}+\bar{\beta}\frac{\partial}{\partial x}\bigl(x\bar{W}\bigl)+\hat{\mathcal{M}}\frac{\partial}{\partial x}\bar{W} +x\hat{\mathcal{M}}_3\bar{W}\nonumber\\
&\quad-k_1x\hat{\mathcal{M}}_4\bar{W}+\pazocal{L}_{\mathrm{2LS}}\bar{W}+\hat{\mathcal{M}}^{sc}\bar{W},
\end{align}
\noindent
where the superoperator $\hat{\mathcal{M}}=\hat{\mathcal{M}}_2+k_1\hat{\mathcal{M}}_1-k_2\hat{\mathcal{M}}_4$, and the coefficients are $\bar{\alpha}=k_B T \omega/\gamma \hbar$, $\bar{\beta}=\omega^2/2\gamma$, $k_1=\omega/2\gamma$, and $k_2=2k_BT/\gamma\hbar$. The spin-coupling term $\hat{\mathcal{M}}^{sc}\bar{W}$ is given explicitly in the Appendix~\ref{spin_coupling}.
The master equation~(\ref{finmee}) describes the OQBM, and it has the same form as the master equation introduced by Bauer~\textit{et al.} (see equation (2) in~\cite{bauer2013open} and equation (28) in~\cite{bauer2014open}).
The diffusive term
\begin{equation}
\bar{\alpha}\frac{\partial^2 }{\partial x^2}\bar{W}+\bar{\beta}\frac{\partial}{\partial x}\bigl(x\bar{W}\bigl),
\end{equation}
\noindent
describes the propagation of the Brownian particle. The Lindblad and spin-coupling terms
\begin{equation}
\pazocal{L}_{\mathrm{2LS}}\bar{W}+\hat{\mathcal{M}}^{sc}\bar{W}
\end{equation}
\noindent
describes the dynamics of the internal degree of freedom of the Brownian particle. The quantum coin term
\begin{equation}\label{cointerm}
   \hat{\mathcal{M}}\frac{\partial}{\partial x}\bar{W}+x\hat{\mathcal{M}}_3\bar{W}-k_1x\hat{\mathcal{M}}_4\bar{W},
\end{equation}
\noindent
describes the interaction between the quantum Brownian particle's external and internal degrees of freedom. The quantum Brownian motion becomes open in the presence of this term~(\ref{cointerm}), which acts like a `decision-making' term and influences the direction of propagation of the Brownian particle. Equation~(\ref{finmee}) concludes the derivation of the OQBM. 

In summary, the master equation~(\ref{finmee}) is valid for $\omega,\Omega\ll \gamma\ll  \Lambda \ll k_B T/\hbar$, although we cannot rule out the possibility that additional constraints may be required.
In Sect.~\ref{part_3}, we illustrate the derivation by studying the OQBM dynamics, and in Sect.~\ref{part_4}, we derive the equations for the $n$th moments of the position distribution of the OQBM walker.

%%%%%%%%%%%%%%%%%%%%%%%%%%%%%%%%%%%%%%%%%%%%%%%%%%%%%%%%%%%%%
\section{Numerical examples of OQBM dynamics}\label{part_3}
\begin{figure}[htbp]
\centering 
{%
\includegraphics[width=8.6cm]{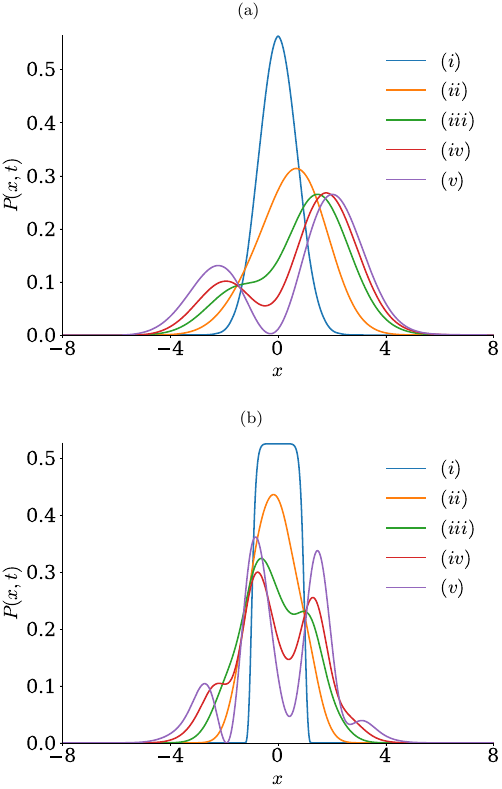}%
}
\caption{The position probability distribution of the open quantum Brownian particle for different times. The initial distribution is given by Eq.~(\ref{fn_1}), with $L=8$. Curves ($i$) through ($v$) correspond to times 0, 50, 100, 150, and 200, respectively. For subplot (a), the initial distribution is chosen to be Gaussian (with $k=2$), and the internal degree of freedom parameters are set to $\theta=\pi/6$, and $\phi=\pi/2$. Other parameters are $\bar{\alpha}=8.0\times10^{-3}$, $\bar{\beta}=10^{-5}$, $\bar{\beta}_1=1.55\times10^{-2}$, $\bar{\beta}_2=9.5\times10^{-3}$, $\bar{\beta}_3=1.33\times10^{-2}$, $\bar{\lambda}_1=-10^{-4}$, $\bar{\lambda}_2=10^{-3}$, $k_1=10^{-3}$, $k_2=1.4$, $k_3=2.0$, and $k_4=0.1$. For subplot (b), the initial distribution is chosen to be non-Gaussian (with $k=10$), and the internal degree of freedom parameters are set to $\theta=\pi/6$, and $\phi=0$. The remaining parameters are  $\bar{\alpha}=2.0\times10^{-3}$, $\bar{\beta}=10^{-5}$, $\bar{\beta}_1=1.3\times10^{-2}$, $\bar{\beta}_2=8.0\times10^{-3}$, $\bar{\beta}_3=10^{-2}$, $\bar{\lambda}_1=-10^{-4}$, $\bar{\lambda}_2=1.03\times10^{-2}$, $k_1=10^{-3}$, $k_2=1.9$, $k_3=5.0$, and $k_4=1.0$.}
\label{example1}
\end{figure}

%%%%%%%%%%%%%%%%%%%%%%%%%%%%%%%%%%%%%%%%%%%%%%%%%%%%%%%%%%%%%
The reduced Wigner function $\bar{W}(x,t)$ of the open quantum Brownian particle can be written in the matrix form as
\begin{equation}\label{densitym}
  \bar{W}(x,t) = \begin{pmatrix}
W_{1,1}(x,t) & W_{1,2}(x,t) \\
W_{2,1}(x,t) & W_{2,2}(x,t)
\end{pmatrix},
\end{equation}
\noindent
where $W_{1,1}(x,t)$ and $W_{2,2}(x,t)$ denote the probability density of finding the open quantum Brownian particle at the position, $x$, at time, $t$, and the off-diagonal elements $W_{1,2}(x,t)=(W_{2,1}(x,t))^*$ represent the coherences.
Using the representation in Eq.~(\ref{densitym}), the master equation~(\ref{finmee}) can be written as a system of partial differential equations, which reads:
\begin{align}\label{systemss}
    \frac{\partial}{\partial t}W_+ &=  \biggl\{\bar{\alpha} \frac{\partial^2}{\partial x^2} +\bar{\beta}\frac{\partial}{\partial x}x\biggl\}W_++\delta_1\frac{\partial}{\partial x}C_R+\delta_2\frac{\partial}{\partial x}C_I,\nonumber\\
     \frac{\partial}{\partial t}W_- &=\biggl\{\bar{\alpha}\frac{\partial^2}{\partial x^2}+\bar{\beta}\frac{\partial}{\partial x}x-\delta_6\biggl\}W_-+\biggl\{\delta_4\frac{\partial}{\partial x} -\delta_1 x\biggl\}C_R\nonumber\\
     &\quad+\biggl\{\delta_5\frac{\partial}{\partial x}+\delta_2 x\biggl\}C_I+\delta_3W_+,\nonumber\\
     \frac{\partial}{\partial t}C_R&=\biggl\{\bar{\alpha}\frac{\partial^2}{\partial x^2}+\bar{\beta}\frac{\partial}{\partial x}x-\delta_7\biggl\}C_R+\frac{\delta_1}{4}\frac{\partial}{\partial x}W_++\delta_8 C_I\nonumber\\
     &\quad-\frac{1}{4}\biggl\{\delta_4\frac{\partial}{\partial x}-\delta_1x\biggl\}W_-,\nonumber\\
     \frac{\partial}{\partial t} C_I&=\biggl\{\bar{\alpha}\frac{\partial^2}{\partial x^2}+\bar{\beta}\frac{\partial}{\partial x}x+\delta_{10}\biggl\}C_I+\frac{\delta_2}{4}\frac{\partial}{\partial x}W_+\nonumber\\
     &\quad-\frac{1}{4}\biggl\{\delta_5\frac{\partial}{\partial x}+\delta_2x\biggl\}W_--\delta_{9}C_R,
\end{align}
\noindent
where $W_{\pm}=W_{1,1}(x,t)\pm W_{2,2}(x,t)$, $C_R=\mathrm{Re}(W_{1,2}(x,t))$, and $C_I=\mathrm{Im}(W_{1,2}(x,t))$. For notational convenience, the combinations of coefficients introduced in Eq.~(\ref{systemss}) are defined as follows:
\begin{align}
\delta_1 &= \bar{\beta}_2-2k_1 \bar{\beta}_3, \hspace{2mm} \delta_2=2\bar{\beta}_3+k_1\bar{\beta}_2, \hspace{2mm} \delta_3= \frac{k_4}{4}\bar{\beta}_2^2-k_4\bar{\beta}_3^2,\nonumber\\
\delta_4&=2k_2\bar{\beta}_3-2\bar{\beta}_2-4\bar{\beta}_1,\hspace{2mm}\delta_5=k_2\bar{\beta}_2-2k_1\bar{\beta}_2-4k_1\bar{\beta}_1,\nonumber\\
\delta_6 &=\frac{5}{4}k_3\bar{\beta}_3^2-\frac{k_4}{2}\bar{\beta}_2\bar{\beta}_3-k_4\bar{\beta}_1\bar{\beta}_3+2\bar{\lambda}_2,\hspace{2mm}
\delta_7=k_3\bar{\beta}_3^2+\bar{\lambda}_2,\nonumber\\
\delta_8&=2k_4\bar{\beta}_1\bar{\beta}_3-\frac{k_3}{2}\bar{\beta}_2\bar{\beta}_3+k_4\bar{\beta}_2\bar{\beta}_3-2\bar{\lambda}_1,\nonumber\\
\delta_9&=\frac{k_3}{2}\bar{\beta}_2\bar{\beta}_3-2\bar{\lambda}_1,
\nonumber\\
\delta_{10}&=\frac{k_4}{2}\bar{\beta}_2\bar{\beta}_3-\frac{k_3}{2}\bar{\beta}_3^2+k_4\bar{\beta}_1\bar{\beta}_3-\bar{\lambda}_2.
\end{align}
To investigate the OQBM dynamics, we numerically integrate the system of partial differential equations~(\ref{systemss}).
For demonstration purposes, we examine both the dynamics of the Gaussian and non-Gaussian initial position distributions of the quantum Brownian particle, and we assume that the internal degree of freedom is initially in a pure state.
The probability $P(x,t) = W_+(x,t)$ of finding the open quantum Brownian particle at a specific position, $x$, after time, $t$, is displayed in Fig.~\ref{example1}.
For this example, we chose the following function as the initial distribution of the open quantum Brownian
particle
\begin{align}\label{fn_1}
  \bar{W}_k(x,0) = \frac{1}{2I_k}e^{-|x|^k}\otimes\begin{pmatrix}
2\cos^2\theta & \sin 2\theta e^{-i\phi} \\
\sin 2\theta e^{i\phi} & 2\sin^2\theta
\end{pmatrix},	
\end{align}
\noindent
where $I_k=\int_{-L}^{+L} dx\, e^{-|x|^k}$, $L$ denotes the location of the confining boundaries for the walker, $\theta\in [0,\pi)$, $\phi\in [0,2\pi)$, and $k>0$.
As illustrated in Fig.~\ref{example1}(a), for the case of $k=2$, it is evident that the initial Gaussian distribution for a chosen set of system-bath parameters separates into two Gaussian distributions after sufficient time, e.g., $t>100$. The internal degree of freedom (qubit) transfers the coherent component to the spatial part, causing the initial Gaussian distribution with zero expectation value $\langle x \rangle = 0$ to shift to a distribution with a nonzero expectation value after the interaction of the qubit and external degree of freedom. Figure~\ref{example1}(b), for the case of $k=10$, demonstrates that even with an explicitly non-Gaussian initial distribution, the position probability distribution of the open Brownian particle becomes Gaussian after sufficient time, e.g., $t\ge50$ for a chosen set of system-bath parameters. 
From both examples, we note that the direction of spread and the limiting distribution asymmetry depend on the qubit's initial state ($\theta$ and $\phi$) and the chosen set of parameters.
The number of peaks that appear at times $t>100$ is not limited to two peaks; see Fig.~\ref{example1}(b), curve~($v$). 
One can generate more than two peaks by adjusting the system-bath parameters and the initial state of the qubit ($\theta$ and $\phi$). 
A qubit initialized with zero coherences leads to a symmetric limiting distribution, whereas nonzero coherences in the initial state result in an asymmetric limiting distribution. Meanwhile, OQWs~\cite{ATTAL20121545,attal2012open} converge to a distribution limited to only two Gaussian profiles. OQBM, on the other hand, are characterized by multiple Gaussian profiles and broader spreading, highlighting different transport dynamics as demonstrated in Fig.~\ref{example1}(b).

\begin{figure}[htbp]
\centering{%
  \includegraphics[width=8.6cm]{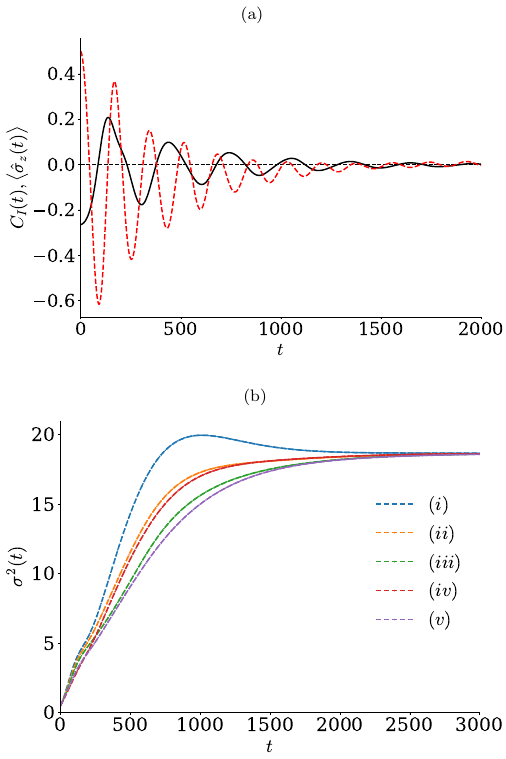}%
}
\caption{OQBM dynamics. The initial distribution is given by Eq.~(\ref{fn_1}), with $k=2$. In subplot (a), we plot the time evolution of the imaginary part of the off-diagonal element $\bigl(C_I (t) = {\tr}_x\bigl[C_I(x,t)\bigl]\bigl)$ (solid curve) and the expectation value of $\langle \hat{\sigma}_z (t) \rangle$ (dashed curve) of the open quantum Brownian particle. The parameters specifying the internal degree of freedom are chosen as $\theta=\pi/6$, and $\phi=\pi/4$, with the spatial domain set to $L=8$. Other parameters are $\bar{\alpha}=10^{-3}$, $\bar{\beta}=10^{-5}$, $\bar{\beta}_1=2.8\times10^{-2}$, $\bar{\beta}_2=5.0\times10^{-4}$, $\bar{\beta}_3=2.13\times10^{-2}$, $\bar{\lambda}_1=-10^{-4}$, $\bar{\lambda}_2=5.0\times10^{-4}$, $k_1=10^{-3}$, $k_2=2.5$, $k_3=3.0$, and $k_4=0.1$. Subplot (b) shows the variance $\sigma^2(t)$ as a function of time for different initial internal states of the OQBM walker, with the spatial domain set to $L=30$. Curves ($i$)-($v$) correspond to the parameter set $\bar{\alpha}=10^{-2}$, $\bar{\beta}=10^{-3}$, $\bar{\beta}_1=1.5\times10^{-2}$, $\bar{\beta}_2=10^{-2}$, $\bar{\beta}_3=1.2\times10^{-2}$, $\bar{\lambda}_1=-10^{-4}$, $\bar{\lambda}_2=8.0\times10^{-3}$, $k_1=10^{-3}$, $k_2=1.0$, $k_3=5.0$, and $k_4=0.1$. The initial internal states are specified by the parameters $(\theta,\phi)=(0,0)$, $(\pi/6,0)$, $(\pi/4,0)$, $(\pi/4,\pi/2)$, and $(\pi/3,\pi/4)$ for curves $(i)$–$(v)$, respectively.}
\label{example2}
\end{figure}

Further, we investigate the dynamics of the internal degree of freedom; the imaginary part of the off-diagonal element of the OQBM density matrix $\bigl(C_I (t) = \tr_x\bigl[C_I(x,t)\bigl]\bigl)$ and the population imbalance in the internal degree of freedom of the open quantum Brownian particle $\bigl(\langle \hat{\sigma}_z (t) \rangle=\tr\bigl[\bar{W}(x,t)\hat{\sigma}_z\bigl]\bigl)$.  As illustrated in Fig.~\ref{example2}(a), both $C_I(t)$ and $\langle \hat{\sigma}_z(t) \rangle$ exhibit oscillatory dynamics during the evolution, but due to interaction with the bath, both quantities decay to zero over time.

In Fig.~\ref{example2}(b), we plot the time-dependent variance $\sigma^2(t)$ in the position of the OQBM walker for a fixed set of parameters and different initial internal states, with the initial spatial distribution chosen to be Gaussian~(\ref{fn_1}) (with $k=2$, and $L=30$). In all cases, the variance follows a power-law scaling, $\sigma^2(t) \sim t^{\nu}$, over the time interval $10\lesssim t \lesssim 900$. A fit in this interval yields $\nu \approx 1$, indicating the diffusive character of the evolution. For later times, $900\lesssim t \lesssim 1500$, deviations from the power-law behavior become apparent as the dynamics enter a crossover regime. In this interval, the walker begins to feel the confining potential walls, and the transport is no longer diffusive. This behavior is similar to the Ornstein--Uhlenbeck (OU) process, where the presence of a confining potential prevents indefinite spreading and causes the variance to saturate at long times~\cite{gardiner1985handbook}. Curve $(i)$ shows the fastest initial dynamics, while curves $(ii)$ and $(iv)$ show the moderate initial dynamics, with curves $(iii)$ and $(v)$ exhibiting slow behavior. Although we initialized the OQBM walker with different initial internal states parametrized by $\theta$ and $\phi$, these conditions modify the transient spreading rate and the relaxation timescale, with a clear dependence of the transient dynamics on the choice of initial internal state.

\begin{figure}[htbp]
\centering 
{%
\includegraphics[width=8.6cm]{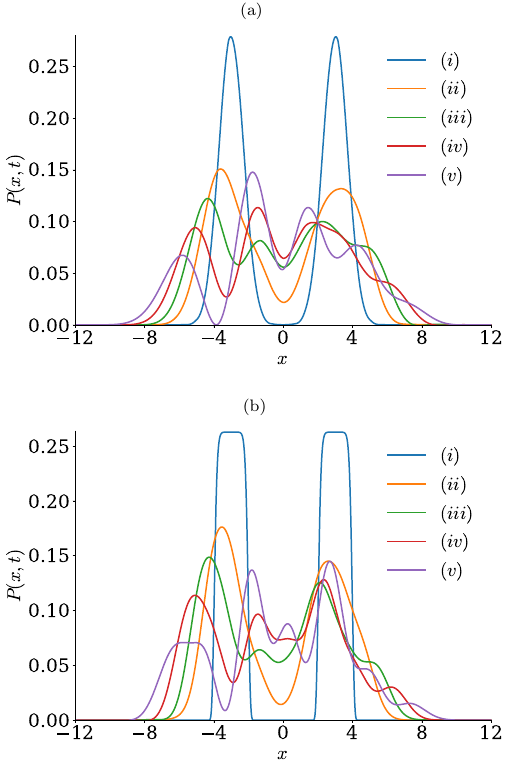}%
}
\caption{The position probability distribution of the open quantum Brownian particle at different times. The initial distribution is given by Eq.~(\ref{fn_2}), with $L=12$. Curves ($i$) through ($v$) correspond to times 0, 50, 100, 150, and 200, respectively. For subplot (a), the initial distribution is given by a Gaussian distribution (with $k=2$), and the internal degree of freedom parameters are set to $\theta=\pi/10$, and $\phi=\pi/6$. Other parameters are $\bar{\alpha}=8.0\times10^{-3}$, $\bar{\beta}=10^{-5}$, $\bar{\beta}_1=2.7\times10^{-2}$, $\bar{\beta}_2=1.3\times10^{-2}$, $\bar{\beta}_3=1.43\times10^{-2}$, $\bar{\lambda}_1=-10^{-4}$, $\bar{\lambda}_2=1.73\times10^{-2}$, $k_1=10^{-3}$, $k_2=1.8$, $k_3=5.0$, and $k_4=0.8$. For subplot (b), the initial distribution is given by a non-Gaussian distribution (with $k=10$), and the internal degree of freedom parameters are set to $\theta=\pi/6$, and $\phi=0$. The remaining parameters are $\bar{\alpha}=8\times10^{-3}$, $\bar{\beta}=10^{-5}$, $\bar{\beta}_1=2.5\times10^{-2}$, $\bar{\beta}_2=1.2\times10^{-2}$, $\bar{\beta}_3=1.3\times10^{-2}$, $\bar{\lambda}_1=-10^{-4}$, $\bar{\lambda}_2=1.7\times10^{-2}$, $k_1=10^{-3}$, $k_2=1.8$, $k_3=5.0$, and $k_4=0.8$.}
\label{example3}
\end{figure}

The examples discussed so far exhibit probability distributions similar to those in Ref.~\cite{sinayskiy2015microscopicbrown}. However, unlike in~\cite{sinayskiy2015microscopicbrown}, our OQBM walker does not propagate far to the left or right because the Brownian particle is confined in a harmonic potential, which restricts the particle's motion.
To generate more interesting dynamics, we now choose a factorized initial state with a bimodal spatial density defined as
\begin{align}\label{fn_2}
  \bar{W}_k(x,0) = \frac{1}{2A_k} f_k(x) \otimes
  \begin{pmatrix}
2\cos^2\theta & \sin 2\theta e^{-i\phi} \\
\sin 2\theta e^{i\phi} & 2\sin^2\theta
\end{pmatrix},	
\end{align}
where,
\begin{equation}
    f_k(x) = e^{-|x+3|^k}+e^{-|x-3|^k}.
\end{equation}
\noindent
Here, $A_k=\int_{-L}^{+L} dx\, f_k(x)$, with $L$, $\theta$, $\phi$ and $k $ as defined previously.
To demonstrate the dynamics of this OQBM, again, we consider Gaussian ($k=2$) and non-Gaussian ($k=10$) initial distributions of the quantum Brownian particle. 
Figure~\ref{example3} shows the probability of finding the open quantum Brownian particle at different moments of time. 
In Fig.~\ref{example3}(a), the initially separated peaks broaden and
overlap at time $t=50$, and the third Gaussian-like profile is formed at time $t\ge100$. At later times, e.g., $t=200$, the OQBM walker's position probability distribution consists of four peaks propagating to the left.
In Fig.~\ref{example3}(b), we chose a non-Gaussian initial distribution, and as in the previous example (see Fig.~\ref{example1}(b)), the probability distribution of finding the open quantum Brownian particle at a position, $x$, after time, $t$, becomes a Gaussian after sufficient time, e.g., $t\ge50$. At time $t=200$, one observes that the initial two non-Gaussian distributions split into six Gaussian-like profiles. For these examples (Fig.~\ref{example3}) and Fig.~\ref{example1}, the asymmetric propagation to the left and right is associated with the presence of coherences in the quantum internal degree of freedom, determined by $\theta$ and $\phi$.

\begin{figure}[htbp]
\centering 
{%
\includegraphics[width=8.6cm]{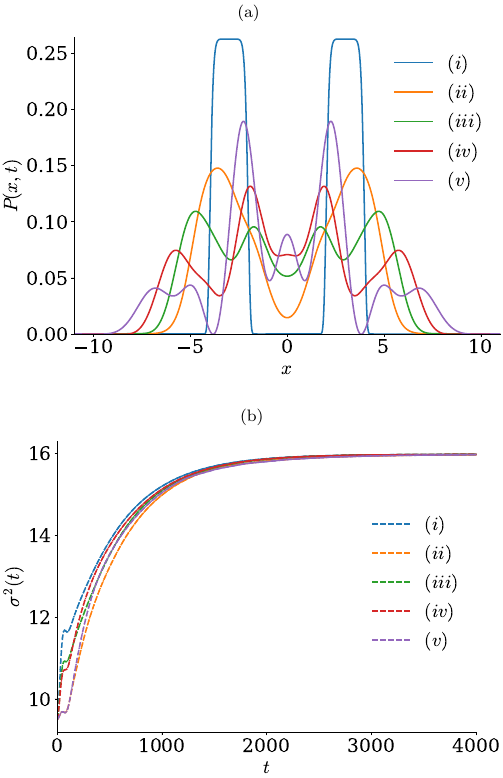}%
}
\caption{OQBM dynamics. Subplot (a) shows the position probability distribution of the open quantum Brownian particle at different times. The initial position distribution is given by Eq.~(\ref{fn_2}) with $k=10$ and $L=11$, while the parameters characterizing the internal degree of freedom are chosen as $\theta=0$ and $\phi=0$. Curves ($i$) through ($v$) correspond to times $0$, $50$, $100$, $150$, and $200$, respectively. The parameters are set to $\bar{\alpha}=8.0\times10^{-3}$, $\bar{\beta}=10^{-5}$, $\bar{\beta}_1=2.5\times10^{-2}$, $\bar{\beta}_2=1.2\times10^{-2}$, $\bar{\beta}_3=1.3\times10^{-2}$, $\bar{\lambda}_1=-10^{-4}$, $\bar{\lambda}_2=1.5\times10^{-2}$, $k_1=10^{-3}$, $k_2=1.82$, $k_3=5.0$, and $k_4=0.8$. Subplot (b) shows the variance $\sigma^2(t)$ as a function of time for different initial internal states of the OQBM walker, with the spatial domain set to $L=40$. Curves ($i$)-($v$) correspond to the parameter set $k=2$, $\bar{\alpha}=9.5\times10^{-3}$, $\bar{\beta}=10^{-3}$, $\bar{\beta}_1=2.6\times10^{-2}$, $\bar{\beta}_2=1.1\times10^{-2}$, $\bar{\beta}_3=1.6\times10^{-2}$, $\bar{\lambda}_1=-5.0\times10^{-4}$, $\bar{\lambda}_2=1.5\times10^{-2}$, $k_1=10^{-3}$, $k_2=1.8$, $k_3=3.0$, and $k_4=0.1$. The initial internal states are specified by the parameters $(\theta,\phi)=(0,0)$, $(\pi/4,\pi)$, $(\pi/7,\pi)$, $(\pi/6,\pi/6)$, and $(\pi/4,0)$ for curves $(i)$–$(v)$, respectively.}
\label{example4}
\end{figure}

Figure~\ref{example4}(a) shows the probability distribution of the open quantum Brownian particle for various system-bath parameters. Initially, the probability distribution consists of two non-Gaussian ($k=10$) decoupled profiles~(\ref{fn_2}). At time $t=50$, these profiles become Gaussian and begin to merge, and by time $t=100$, four Gaussian-like peaks emerge, and by time $t=150$, these peaks are fully developed. At time $t=200$, the distribution evolves into seven Gaussian-like peaks; a moderate central peak located at $x=0$, two large symmetric peaks near the center at $x\approx \pm 2$, and four moderately sized peaks at $x\approx \pm 5$ and $x\approx \pm 7$. The symmetric dynamics of the distribution result from the absence of initial coherences in the internal degree of freedom, corresponding to the parameter choice, $\theta=0$ and $\phi=0$.

Figure~\ref{example4}(b) shows the time-dependent variance $\sigma^2(t)$ in the position of the OQBM walker for a fixed set of parameters and different initial internal states. The initial position distribution is chosen to be Gaussian ($k=2$) for a decoupled initial state~(\ref{fn_2}). As shown in Fig.~\ref{example4}(b), the initial differences between the curves ($i$)-($v$) arise from the choice of the initial internal state, which influences the OQBM walker's transport properties through the degree of quantum coherence and population imbalance. As seen from Fig.~\ref{example4}(b), a transition in the dynamics is observed before the onset of the steady state. For example, over an intermediate time interval, the variance follows a power-law scaling, $\sigma^2(t) \sim t^{\nu}$. However, since the earlier times $t \lesssim 100$ show instabilities in the variance, we exclude that interval and fit the data over the time interval $200 \lesssim t \lesssim 1000$, yielding $\nu <1$, which indicates the sub-diffusive character of the evolution. At longer times, $t \gtrsim 1000$, the dynamics enter a crossover regime, during which the variance gradually approaches a nearly constant value and eventually reaches a steady state, indicating that the OQBM walker spreads more slowly due to the confining potential.

Our results demonstrate that the open quantum Brownian particle can propagate in both directions at distinct speeds and spreading rates for a particular choice of system-bath parameters. The symmetry or asymmetry of the distribution depends on the initial coherence of the internal degree of freedom. As illustrated in Fig.~\ref{example4}(a), when the initial coherences of the internal degree of freedom are zero, the distribution remains symmetric; when coherences are present, as in Fig.~\ref{example1}(a)-(b) and Fig.~\ref{example3}(a)-(b), the distribution becomes asymmetric. In all examples, at time $t\approx200$, multiple Gaussian-like peaks are visible. Lastly, the position probability distribution becomes Gaussian at time $t\approx 50$, even for non-Gaussian initial distributions.

%%%%%%%%%%%%%%%%%%%%%%%%%%%%%%%%%%%%%%%%
\section{Moments of the position distribution}\label{part_4}
\begin{figure}[htbp]
\centering 
{%
\includegraphics[width=8.6cm]{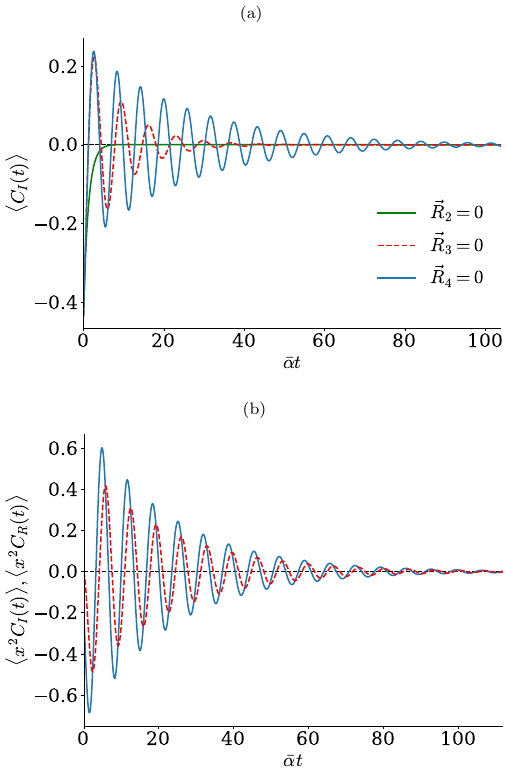}%
}
\caption{Time evolution of the coherence moments of the OQBM density matrix as a function of dimensionless time $\bar{\alpha}t$. The initial distribution is given by Eq.~(\ref{iniit}). Subplot (a) shows the zeroth-order coherence moment $\langle C_I(t)\rangle$ for different truncations of the moment hierarchy, $n_\mathrm{cut}=2$, $3$, and $4$, as indicated in the legend. The internal-state parameters are chosen as $\theta=\pi/4$ and $\phi=\pi/3$. Other parameters are $\bar{\alpha}=2.6\times10^{-3}$, $\bar{\beta}=10^{-4}$, $\bar{\beta}_1=5.5\times10^{-3}$, $\bar{\beta}_2=8.0\times10^{-4}$, $\bar{\beta}_3=3.0\times10^{-3}$, $\bar{\lambda}_1=-6.0\times10^{-4}$, $\bar{\lambda}_2=3.0\times10^{-3}$, $k_1=10^{-3}$, $k_2=4.3$, $k_3=4.5$, and $k_4=1.0$.
Subplot (b) shows second-order moments with respect to the real part $\langle x^2 C_R(t)\rangle$ (dashed curve) and the imaginary part $\langle x^2 C_I(t)\rangle$ (solid curve) of the OQBM density matrix. The internal-state parameters are chosen as $\theta=\pi/6$ and $\phi=\pi/2$, and the moment hierarchy is truncated at $n_\mathrm{cut}=4$. Other parameters are $\bar{\alpha}=2.8\times10^{-3}$, $\bar{\beta}=10^{-4}$, $\bar{\beta}_1=5.0\times10^{-3}$, $\bar{\beta}_2=8.0\times10^{-4}$, $\bar{\beta}_3=3.0\times10^{-3}$, $\bar{\lambda}_1=-5.0\times10^{-4}$, $\bar{\lambda}_2=3.0\times10^{-3}$, $k_1=10^{-3}$, $k_2=4.2$, $k_3=8.0$, and $k_4=1.0$.}
\label{momentss_n}
\end{figure}

%%%%%%%%%%%%%%%%%%%%%%%%%%%%%%%%%%%%%%%%
To obtain a quantitative characterization of the OQBM walker, Eq.~(\ref{systemss}) is used to derive explicit equations of motion for the $n$th moments of the position distribution. We denote the $n$th moment by
\begin{equation}\label{momentss}
    \bigl\langle x^n W(x,t)\bigl\rangle = \int_{-\infty}^{+\infty}  dx\, x^n W(x, t),
\end{equation}
\noindent
where $W(x,t) \equiv \{W_+,W_-,C_R, C_I\}$. By direct substitution of Eq.~(\ref{momentss}) into Eq.~(\ref{systemss}), we obtain an infinite hierarchy of coupled linear ordinary differential equations for the moments:
\begin{align}\label{syst_pde}
    \frac{d}{dt} \bigl\langle x^n W_+\bigl\rangle &= \bar{\alpha} n (n-1) \bigl\langle x^{n-2}W_+ \bigl\rangle-\bar{\beta} n \bigl\langle x^nW_+ \bigl\rangle\nonumber\\
    &-\delta_1n\bigl\langle x^{n-1} C_R\bigl\rangle-\delta_2n\bigl\langle x^{n-1} C_I\bigl\rangle,\nonumber\\
    \frac{d}{dt}\bigl\langle x^n W_-\bigl\rangle&=\bar{\alpha}n(n-1)\bigl\langle x^{n-2} W_-\bigl\rangle-\bigl(\bar{\beta}n+\delta_6\bigl)\bigl\langle x^n W_-\bigl\rangle\nonumber\\
    &+\delta_3\bigl\langle x^n W_+\bigl\rangle-\delta_4n\bigl\langle x^{n-1} C_R\bigl\rangle-\delta_5n\bigl\langle x^{n-1}C_I\bigl\rangle\nonumber\\
    &-\delta_1\bigl\langle x^{n+1}C_R\bigl\rangle+\delta_2\bigl\langle x^{n+1}C_I\bigl\rangle,\nonumber\\
    \frac{d}{dt}\bigl\langle x^n C_R\bigl\rangle&=\bar{\alpha}n(n-1)\bigl\langle x^{n-2}C_R\bigl\rangle-\bigl(\bar{\beta}n+\delta_7\bigl)\bigl\langle x^n C_R\bigl\rangle\nonumber\\
    &+\frac{\delta_4}{4}n\bigl\langle x^{n-1} W_-\bigl\rangle+\frac{\delta_1}{4}\bigl\langle x^{n+1} W_-\bigl\rangle+\delta_8\bigl\langle x^n C_I\bigl\rangle\nonumber\\&-\frac{\delta_1}{4}n\bigl\langle x^{n -1}W_+\bigl\rangle,\nonumber\\
    \frac{d}{dt}\bigl\langle x^n C_I\bigl\rangle&=\bar{\alpha}n(n-1)\bigl\langle x^{n-2}C_I\bigl\rangle-\bigl(\bar{\beta}n-\delta_{10}\bigl)\bigl\langle x^nC_I\bigl\rangle\nonumber\\
    &+\frac{\delta_5}{4}n\bigl\langle x^{n -1}W_-\bigl\rangle-\frac{\delta_2}{4}\bigl\langle x^{n+1}W_-\bigl\rangle-\delta_9\bigl\langle x^nC_R\bigl\rangle\nonumber\\
    &-\frac{\delta_2}{4}n\bigl\langle x^{n -1}W_+\bigl\rangle.
\end{align}
\noindent
%This is a linear infinite-dimensional system of coupled differential equations for the moments. 
For ease of numerical integration, we rewrite the system~(\ref{syst_pde}) in the following form:
\begin{align}\label{systemsf}
    \frac{d}{dt}\vec{R}_n = \ &\hat{M}_n \vec{R}_n + \hat{A}_n \vec{R}_{n-1}+\hat{B}_n\vec{R}_{n-2}+\hat{C}\vec{R}_{n+1},
\end{align}
\noindent
where $ \vec{R}_n$,  $ \vec{R}_{n-1}$, $ \vec{R}_{n-2}$, and $ \vec{R}_{n+1}$ denote column vectors containing the moment components at orders $n$, $n-1$, $n-2$, and $n+1$, respectively,  defined as
\begin{equation}
    \vec{R}_{n + i} = \begin{pmatrix}
        \langle x^{n+i} W_+\rangle  \\
        \langle x^{n+i} W_-\rangle  \\
        \langle x^{n+i} C_R\rangle \\
        \langle x^{n+i} C_I\rangle 
    \end{pmatrix}, \hspace{4mm} i=0,\pm 1,-2.
\end{equation}
\noindent
The operators $\hat{M}_n$, $\hat{A}_n$, $\hat{B}_n$ are four-by-four matrices of parameters that depend  on the index $n$, while $\hat{C}$ is a four-by-four matrix independent of $n$, defined as:
\begin{align}
    \hat{M}_n& =  \begin{pmatrix}
       -\bar{\beta}n&0&0&0 \\
        \delta_3&-(\bar{\beta}n+\delta_6)&0&0 \\
        0&0&-(\bar{\beta}n+\delta_7)&\delta_8 \\
         0&0&-\delta_9 &-\bar{\beta}n+\delta_{10}\\
    \end{pmatrix},\nonumber\\
    \hat{A}_n &=  n\begin{pmatrix}
       0&0&-\delta_1&-\delta_2 \\
    0&0&-\delta_4&-\delta_5 \\
        -\frac{1}{4}\delta_1&\frac{1}{4}\delta_4&0&0 \\
         -\frac{1}{4}\delta_2&\frac{1}{4}\delta_5&0&0\\
    \end{pmatrix},\nonumber\\
    \hat{C} &=  \begin{pmatrix}
       0&0&0&0 \\
        0&0&-\delta_1&\delta_2 \\
        0&\frac{1}{4}\delta_1&0&0 \\
         0&-\frac{1}{4}\delta_2&0&0\\
    \end{pmatrix},\hspace{1mm} \hat{B}_n = \bar{\alpha}n(n-1)\hat{I}_{4\times4}. 
\end{align}
\noindent
The system~(\ref{systemsf}) is solved numerically for the moments by imposing the truncation condition $\vec{R}_{n_\mathrm{cut}}=0$, where $n_\mathrm{cut}$ denotes the number of retained moment orders. The initial spatial distribution is taken to be Gaussian, with the internal degree of freedom assumed to be initially in a pure state:
\begin{equation}\label{iniit}
  W(x,0) = \frac{1}{2\sqrt{\pi}}e^{-x^2}\otimes\begin{pmatrix}
2\cos^2\theta & \sin 2\theta e^{-i\phi} \\
\sin 2\theta e^{i\phi} & 2\sin^2\theta
\end{pmatrix}.
\end{equation}
\noindent
At time $t=0$, Eq.~(\ref{momentss}) shows that the arbitrary initial conditions, expressed as functions of $n$, are given by
\begin{align}\label{initial_conds}
   \bigl\langle x^n W_\pm\bigl\rangle =\ &   \frac{1+(-1)^n}{2\sqrt{\pi}}\Gamma\biggl(\frac{1+n}{2}\biggl)W_\pm,\nonumber\\
      \bigl\langle x^n C_R\bigl\rangle =\ &   \frac{1+(-1)^n}{2\sqrt{\pi}}\Gamma\biggl(\frac{1+n}{2}\biggl)C_R,\nonumber\\
         \bigl\langle x^n C_I\bigl\rangle =\ &   \frac{1+(-1)^n}{2\sqrt{\pi}}\Gamma\biggl(\frac{1+n}{2}\biggl)C_I.
\end{align}
\noindent
For the symmetric Gaussian spatial state, all odd moments vanish initially. The coupled internal state-dependent dynamics may generate nonzero odd moments for $t>0$. Here, the functions $W_\pm$ and  $C_{R, I}$ have the same meaning as defined earlier. 
The derived system~(\ref{syst_pde}) becomes strongly coupled as the moment order $n$ increases, making the system numerically stiff at higher-order moments, particularly when the hierarchy is truncated at $\vec{R}_5=0$. Therefore, our analysis is restricted to $n_\mathrm{cut}\le4$. 
Figure~\ref{momentss_n}(a) shows the zeroth-order coherence moment $\langle C_I(t)\rangle$ as a function of dimensionless time for different truncations, $\vec{R}_2=0$, $\vec{R}_3=0$, and $\vec{R}_4=0$, corresponding to $n_\mathrm{cut}=2$, $3$, and $4$, respectively (as indicated in the legend). For $n_\mathrm{cut}=2$, no oscillations are observed, while weak oscillations emerge for $n_\mathrm{cut}=3$ and become more pronounced for $n_\mathrm{cut}=4$, where they are gradually damped over time. These results demonstrate that the coherence dynamics are sensitive to the truncation order, indicating that higher-order spatial moments play a role in the dynamics of the zeroth-order coherence moment.

Figure~\ref{momentss_n}(b) shows the dynamics of the second-order moments with respect to the imaginary part $\langle x^2 C_I(t)\rangle$ and the real part $\langle x^2 C_R(t)\rangle$ of the OQBM density matrix as a function of dimensionless time for various system-bath parameters. For this example,  the moment hierarchy is truncated at $\vec{R}_4=0$, and the initial state of the internal degree of freedom is specified by the parameters $\theta=\pi/6$ and $\phi=\pi/4$. As illustrated in Fig.~\ref{momentss_n}(b), both $\langle x^2 C_I(t)\rangle$ and $\langle x^2 C_R(t)\rangle$ exhibit initial coherences during the evolution, with different oscillation amplitudes. However, due to the system's interaction with the bath, both quantities decay to zero over time.
These results demonstrate that an adequate truncation order is essential for accurately capturing the coherence dynamics.

\section{Summary and conclusion}\label{part_5}

This paper presents the first generic microscopic derivation of Open Quantum Brownian Motion (OQBM) for a Brownian particle confined in a harmonic potential subject to decoherent interaction with a thermal bath.
Starting from the Hamiltonian of the Brownian particle with a single quantum internal degree of freedom, the bath Hamiltonian, and the system-bath interaction Hamiltonian, we assume a high-temperature limit for the bath and derive the Born--Markov master equation for the reduced density matrix. The resulting master equation is written in phase-space representation using the Wigner function. We proceed with the derivation of OQBM by assuming a high-damping limit. This assumption forces the Brownian particle's momentum variable to thermalize quickly and rapidly reach equilibrium with the bath. In contrast, the Brownian particle's position variable evolves more slowly; as a result, we eliminate the momentum variable adiabatically and derive the master equation for the OQBM. 

We illustrate the derivation by presenting examples of OQBM dynamics with initial Gaussian and non-Gaussian distributions across various system-bath parameters. In all examples, the position probability distribution of finding the open quantum Brownian particle at different moments converges to Gaussian distributions after sufficient time, e.g., $t\ge50$. We noted that the internal degree of freedom initialized with zero coherences leads to a symmetric limiting distribution, whereas non-zero coherences in the internal degree of freedom result in an asymmetric limiting distribution for various parameters.

Unlike the examples in Refs.~\cite{sinayskiy2015microscopicbrown,sinayskiy2017steady}, our OQBM walker's propagation is restricted because the particle is confined in a harmonic potential. The number of Gaussian profiles at the finite simulation time $t\approx 200$ is not limited to two Gaussian distributions as in the case of OQWs~\cite{ATTAL20121545,attal2012open}, highlighting a rich dynamical behavior.

We also investigated the dynamics of the internal degree of freedom and found that both quantities $C_I(t)$ and $\langle \hat{\sigma}_z(t)\rangle$ initially exhibit damped oscillatory dynamics and decay to zero as a function of time.
In addition, we plotted the variance $\sigma^2(t)$ as a function of time at the position of the OQBM walker, and we observed initial diffusive spreading, followed by a crossover to a finite steady-state width.
Furthermore, we derived the equations for the $n$th moments of the OQBM walker's position distribution. Since the system becomes strongly coupled as the order $n$ increases, we restricted our numerical examples to $n_\mathrm{cut}\leq4$. Both examples exhibit damped oscillations, consistent with damped internal-state dynamics generated by the dissipative terms. We found that the oscillations become more pronounced for higher truncation orders.

In this paper, we successfully presented the first
generic microscopic derivation of OQBM using the adiabatic elimination of fast variables method. 
This hybrid quantum-classical master equation has the same structural form as
that proposed by Bauer~\textit{et al.}~\cite{bauer2013open,bauer2014open} and subsequently derived in Refs.~\cite{sinayskiy2015microscopicbrown,sinayskiy2017steady}. However, this approach resulted in a master equation that is not completely positive, consistent with the limitations of the standard Caldeira--Leggett model~\cite{caldeira1983path, CALDEIRA1983374}. This generic OQBM offers various possible generalizations and extensions. 
The derivation of the completely positive OQBM master equation in a generic dissipative case will be given elsewhere~\cite{zungu2026microscopic}.

%The derivation of the completely positive OQBM master equation in a generic dissipative case is the subject of future research.

%\begin{acknowledgments}

%This work is based upon research supported by the National Research Foundation of the Republic of South Africa. The author gratefully acknowledges Dr. Camille Lombard-Latune for valuable comments and careful proofreading of the manuscript.
%\end{acknowledgments}

\vspace{5mm}
\noindent
{\bf Acknowledgments.} 
The authors gratefully acknowledge Dr. Camille Lombard Latune for valuable comments and careful proofreading of the manuscript. This work is based upon research supported by the National Research Foundation (NRF) of the Republic of South Africa. AZ acknowledges support in part by the NRF of South Africa (Grant No. 129457). 

\vspace{5mm}
\noindent 
{\bf Conflict of Interest.} 
The authors declare no conflict of interest.

\vspace{5mm}
\noindent 
{\bf Data Availability Statement.}
No new data were generated or analyzed in support of this research.

\vspace{5mm}
\noindent 
{\bf Keywords.} Open quantum systems, open quantum Brownian motion, adiabatic elimination.

\appendix
\section{Derivation of Eq.~(\ref{finmee})}\label{appendixa}

To derive Eq.~(\ref{finmee}), we evaluate the action of the following operators on $v$:
\begin{align}\label{a1}
    &\mathcal{P}\hat{L}_2 \hat{L}_1^{-1}\hat{L}_2 v, \hspace{2mm} \mathcal{P}\hat{L}_2 \hat{L}_1^{-1} \biggl(p\hat{\mathcal{M}}_4+\hat{\mathcal{M}}_1\frac{\partial}{\partial p}\biggl)v,\nonumber\\
    &\mathcal{P}p\hat{\mathcal{M}}_4\hat{L}_1^{-1}\hat{L}_2v,\hspace{2mm}\text{and}\hspace{2mm} \mathcal{P}p\hat{\mathcal{M}}_4 \hat{L}_1^{-1} \biggl(p\hat{\mathcal{M}}_4+\hat{\mathcal{M}}_1\frac{\partial}{\partial p}\biggl)v.
\end{align}
\noindent
By applying $\hat{L}_2$ to $v$, we obtain
\begin{align}\label{a2}
\hat{L}_2 v=-\biggl(\omega\frac{\partial }{\partial x}+\frac{2u(x)}{\alpha}\biggl)P_1(p)\bar{W}.
\end{align}
\noindent
We now employ the following relations
\begin{align}\label{bookeqg}
P_n(p)= \ &(\pi\alpha)^{-1/2}\exp(-p^2/\alpha)Q_n(p),\nonumber\\
Q_n(p) = \ & (2^n n!)^{-1/2}H_n(p/\sqrt{\alpha}),\nonumber\\
\hat{L}_1P_n(p)=\ & - 2n P_n(p), 
\end{align}
\noindent
and the recursion formula for Hermite polynomials
\begin{align}\label{hermite}
xH_n(x)=\ & \frac{1}{2}H_{n+1}(x)+nH_{n-1}(x),\nonumber\\
\frac{d}{dx}\Bigl[e^{-x^2}H_{n}(x)\Bigl]= \ &-e^{-x^2}H_{n+1}(x),
\end{align}
\noindent
adapted from~\cite{gardiner1985handbook} (see Eqs.~(6.4.57)-(6.4.60)). Using Eqs.~(\ref{a2})-(\ref{bookeqg}), it is straightforward to show that
\begin{align}\label{xgadd}
\hat{L}_1^{-1}\hat{L}_2 v=\frac{1}{2}\biggl(\omega\frac{\partial }{\partial x}&+\frac{2u(x)}{\alpha}\biggl)P_1(p)\bar{W},\nonumber\\
\hat{L}_1^{-1}\biggl(p\hat{\mathcal{M}}_4+\hat{\mathcal{M}}_1\frac{\partial}{\partial p}\biggl)v=&-\frac{1}{2}p\hat{\mathcal{M}}_4P_0(p)\bar{W}\nonumber\\
&-\frac{1}{2}\hat{\mathcal{M}}_1\frac{\partial}{\partial p}P_0(p)\bar{W}. 
\end{align}
\noindent
Applying $\hat{L}_2$ once more to Eq.~(\ref{xgadd}) and using the Hermite polynomial recursion formulas~(\ref{hermite}), we obtain
\begin{align}\label{eqnss}
\hat{L}_2P_1(p)=&-\biggl[\sqrt{\alpha}P_2(p)+\sqrt{\frac{\alpha}{2}}P_0(p)\biggl]\biggl(\omega\frac{\partial}{\partial x}\biggl)\nonumber\\
&-\sqrt{\frac{4}{\alpha}}P_2(p)u(x),\nonumber\\
\mathcal{P}\hat{L}_2\hat{L}_1^{-1}\hat{L}_2 v=&-P_0(p)\biggl[\biggl(\frac{k_B T \omega}{\hbar}\biggl)\frac{\partial^2}{\partial x^2}\bar{W}\nonumber\\
&+\biggl(\frac{\omega^2}{2}\biggl)\frac{\partial}{\partial x}\bigl(x\bar{W}\bigl)\biggl],
\end{align}
and 
\begin{align}\label{eqnssx}
&\mathcal{P}\hat{L}_2\hat{L}_1^{-1}\biggl(p\hat{\mathcal{M}}_4+\hat{\mathcal{M}}_1\frac{\partial}{\partial p}\biggl)v\nonumber\\
&= -P_0(p)\biggl(\frac{\omega}{2}\hat{\mathcal{M}}_1-\frac{k_B T}{\hbar}\hat{\mathcal{M}}_4\biggl)\frac{\partial}{\partial x}\bar{W}.
\end{align}
\noindent
Proceeding analogously for the remaining two operators~(\ref{a1}), we obtain
\begin{align}
\mathcal{P}p\hat{\mathcal{M}}_4\hat{L}_1^{-1}\hat{L}_2v=P_0(p)\hat{\mathcal{M}}_4\biggl(\frac{k_B T}{\hbar}\frac{\partial}{\partial x}+\frac{\omega}{2}x\biggl)\bar{W},
\end{align}
and
\begin{align}\label{a9}
&\mathcal{P}p\hat{\mathcal{M}}_4\hat{L}_1^{-1}\biggl(p\hat{\mathcal{M}}_4+\hat{\mathcal{M}}_1\frac{\partial}{\partial p}\biggl)v\nonumber\\
&=\frac{1}{2}P_0(p)\hat{\mathcal{M}}_4\biggl(\hat{\mathcal{M}}_1-\frac{2k_B T}{\hbar\omega}\hat{\mathcal{M}}_4\biggl)\bar{W}.
\end{align}
\noindent
By using Eqs.~(\ref{eqnss})-(\ref{a9}) and $v=P_0(p)\bar{W}$, Eq.~(\ref{felimp}) becomes
\begin{align}\label{finme}
 & \frac{\partial}{\partial t} \bar{W} = \ \biggl(\frac{k_B T \omega}{\gamma \hbar}\biggl)\frac{\partial^2 }{\partial x^2}\bar{W}+\frac{\omega^2}{2\gamma}\frac{\partial}{\partial x}\bigl(x\bar{W}\bigl)+x\hat{\mathcal{M}}_3\bar{W}\nonumber\\
  &+\biggl(\hat{\mathcal{M}}_2+\frac{\omega}{2\gamma}\hat{\mathcal{M}}_1-\frac{2k_B T}{\gamma\hbar}\hat{\mathcal{M}}_4\biggl)\frac{\partial}{\partial x}\bar{W}+\pazocal{L}_{\mathrm{2LS}}\bar{W}\nonumber\\
  &+\hat{\mathcal{M}}_4\biggl(\frac{k_BT}{\gamma \hbar \omega}\hat{\mathcal{M}}_4-\frac{\hat{\mathcal{M}}_1}{2\gamma}\biggl)\bar{W}-\frac{\omega}{2\gamma}x\hat{\mathcal{M}}_4\bar{W}.
\end{align}
\noindent
We have eliminated the fast variable $p$, which is assumed to relax rapidly for large $\gamma$. 
Equation~(\ref{finme}) can therefore be written in the simplified form:
\begin{align}\label{finme_master_eq}
 \frac{\partial}{\partial t} \bar{W}(x,t) &= \bar{\alpha}\frac{\partial^2 }{\partial x^2}\bar{W}+\bar{\beta}\frac{\partial}{\partial x}\bigl(x\bar{W}\bigl)+\hat{\mathcal{M}}\frac{\partial}{\partial x}\bar{W} +x\hat{\mathcal{M}}_3\bar{W}\nonumber\\
&\quad-k_1x\hat{\mathcal{M}}_4\bar{W}+\pazocal{L}_{\mathrm{2LS}}\bar{W}+\hat{\mathcal{M}}^{sc}\bar{W},
\end{align}
\noindent
where the superoperator $\hat{\mathcal{M}}$ is defined as 
\begin{align}\label{ap_eq1}
    \hat{\mathcal{M}}= \hat{\mathcal{M}}_2+k_1\hat{\mathcal{M}}_1-k_2\hat{\mathcal{M}}_4.
\end{align}
\noindent
The spin-coupling term $\hat{\mathcal{M}}^{sc}\bar{W}$ is given by
\begin{align}\label{ap_eq2}
\hat{\mathcal{M}}^{sc}\bar{W}=k_3\hat{\mathcal{M}}_4^2\bar{W}-k_4\hat{\mathcal{M}}_4\hat{\mathcal{M}}_1\bar{W},
\end{align}
\noindent
where 
\begin{align}\label{appedn1}
\hat{\mathcal{M}}_4^2\bar{W}&=\frac{\bar{\beta}_2^2}{8}\pazocal{L}[{\hat{\sigma}_x,\hat{\sigma}_x}]\bar{W}+\frac{\bar{\beta}_3^2}{2}\pazocal{L}[{\hat{\sigma}_y,\hat{\sigma}_y}]\bar{W}\nonumber\\
    &\quad+\frac{\bar{\beta}_2\bar{\beta}_3}{4}\Bigl(\pazocal{L}[\hat{\sigma}_x,\hat{\sigma}_y]\bar{W}+\pazocal{L}[\hat{\sigma}_y,\hat{\sigma}_x]\bar{W}\Bigl),
\end{align}
\noindent
and
\begin{align}\label{appedn2}
&\hat{\mathcal{M}}_4\hat{\mathcal{M}}_1\bar{W}=A_0\pazocal{L}[\hat{\sigma}_x,\hat{\sigma}_x]\bar{W}+(A_1+iA_2)\pazocal{L}[\hat{\sigma}_x,\hat{\sigma}_y]\bar{W}\nonumber\\&+(A_1-iA_2)\pazocal{L}[\hat{\sigma}_y,\hat{\sigma}_x]\bar{W}+iA_1[\hat{\sigma}_z,\bar{W}].
\end{align}
\noindent
For notational convenience, we define the combinations of coefficients appearing in Eq.~(\ref{appedn2}) as follows:
\begin{align}
    A_0 & =\frac{\bar{\beta}_1\bar{\beta}_2}{2}+\frac{\bar{\beta}_2^2}{4},\hspace{2mm}A_1 =\frac{\bar{\beta}_1\bar{\beta}_3}{2}+\frac{\bar{\beta}_2\bar{\beta}_3}{4},\nonumber\\
    A_2 & =\frac{\bar{\beta}_2^2}{16}-\frac{\bar{\beta}_3^2}{4}.
\end{align}
\noindent
From Eqs.~(\ref{appedn1})-(\ref{appedn2}), the spin-coupling term can be written in the following compact form:
\begin{align}\label{spin_coupling}
&\hat{\mathcal{M}}^{sc}\bar{W}=\biggl(k_3\frac{\bar{\beta}_2^2}{8}-k_4A_0\biggl)\pazocal{L}[\hat{\sigma}_x,\hat{\sigma}_x]\bar{W}+k_3\frac{\bar{\beta}_3^2}{2}\pazocal{L}[\hat{\sigma}_y,\hat{\sigma}_y]\bar{W}\nonumber\\&
+\biggl[\frac{k_3}{4}\bar{\beta}_2\bar{\beta}_3-k_4(A_1+iA_2)\biggl]\pazocal{L}[\hat{\sigma}_x,\hat{\sigma}_y]\bar{W}-ik_4A_1[\hat{\sigma}_z,\bar{W}]\nonumber\\
&+\biggl[\frac{k_3}{4}\bar{\beta}_2\bar{\beta}_3-k_4(A_1-iA_2)\biggl]\pazocal{L}[\hat{\sigma}_y,\hat{\sigma}_x]\bar{W}.
\end{align}
\noindent
The coefficients appearing in Eqs.~(\ref{finme_master_eq})-(\ref{ap_eq2}) are defined as
\begin{align}
    \bar{\alpha} &= \frac{k_B T\omega}{\gamma\hbar}, \hspace{2mm} \bar{\beta}=\frac{\omega^2}{2\gamma}, \hspace{2mm} k_1=\frac{\omega}{2\gamma},
    \nonumber\\
    k_2&=\frac{2k_B T}{\gamma \hbar},\hspace{2mm}
    k_3=\frac{k_B T}{\gamma \hbar\omega}, \hspace{2mm} k_4=\frac{1}{2\gamma}.
\end{align}
\noindent
Equation~(\ref{finme_master_eq}) thus completes the derivation of Eq.~(\ref{finmee}) presented in the main text.
\nocite{*}
\bibliography{oqbm}

@PREAMBLE{
 "\providecommand{\noopsort}[1]{}" 
 # "\providecommand{\singleletter}[1]{#1}%" 
}

@book{breuer2002theory,
  title={{The Theory of Open Quantum Systems}},
  author={Breuer, Heinz-Peter and Petruccione, Francesco},
  year={2002},
  publisher={Oxford University Press, Oxford}
}

@article{smoluchowski1916brownsche,
  title={{{\"U}ber Brownsche Molekularbewegung unter Einwirkung {\"a}u{\ss}erer Kr{\"a}fte und deren Zusammenhang mit der verallgemeinerten Diffusionsgleichung}},
  author={Smoluchowski, Marian V},
  journal={Annalen der Physik},
  volume={353},
  number={24},
  pages={1103--1112},
  year={1916},
  publisher={WILEY-VCH Verlag Leipzig}
}

@book{schlosshauer2007decoherence,
  title={{Decoherence and the Quantum-To-Classical Transition, The Frontiers Collection}},
  author={Schlosshauer, Maximilian A},
  year={2007},
  publisher={Springer}
}

@article{wigner1932quantum,
  title={{On the Quantum Correction For Thermodynamic Equilibrium}},
  author={Wigner, Eugene},
  journal={Phys. Rev.},
  volume={40},
  number={5},
  pages={749},
  year={1932},
  publisher={APS},
doi   ={https://doi.org/10.1103/PhysRev.40.749}
}

@article{hillery1984distribution,
  title={{Distribution functions in physics: Fundamentals}},
  author={Hillery, MOSM and O'Connell, Robert F and Scully, Marlan O and Wigner, Eugene P},
  journal={Phys. Rep.},
  volume={106},
  number={3},
  pages={121--167},
  year={1984},
  publisher={Elsevier},
doi={https://doi.org/10.1016/0370-1573(84)90160-1}
}

@book{gardiner2004quantum,
  title={{Quantum Noise: A Handbook of Markovian and Non-Markovian Quantum Stochastic Methods with Applications to Quantum Optics, Springer Series in Synergetics}},
  author={Gardiner, Crispin and Zoller, Peter},
  year={2004},
  publisher={Springer, Berlin}
}

@book{gardiner1985handbook,
  title     = {{Handbook of Stochastic Methods for Physics, Chemistry and the Natural Sciences}},
  author    = {Gardiner, Crispin W.},
  edition   = {3},
  year      = {2004},
  publisher = {Springer},
  address   = {Berlin, Heidelberg}
}

@article{bauer2013open,
  title = {{Open quantum random walks: Bistability on pure states and ballistically induced diffusion}},
  author = {Bauer, Michel and Bernard, Denis and Tilloy, Antoine},
  journal = {Phys. Rev. A.},
  volume = {88},
  issue = {6},
  pages = {062340},
  numpages = {6},
  year = {2013},
  month = {Dec},
  publisher = {American Physical Society},
  doi = {10.1103/PhysRevA.88.062340},
  url = {https://link.aps.org/doi/10.1103/PhysRevA.88.062340}
}

@article{bauer2014open,
doi = {10.1088/1742-5468/2014/09/P09001},
url = {https://dx.doi.org/10.1088/1742-5468/2014/09/P09001},
year = {2014},
month = {sep},
publisher = {IOP Publishing and SISSA},
volume = {2014},
number = {9},
pages = {P09001},
author = {Michel Bauer and Denis Bernard and Antoine Tilloy},
title = {{The open quantum Brownian motions}},
journal = {J. Stat. Mech.},

}

@article{caldeira1983path,
title = {{Path integral approach to quantum Brownian motion}},
journal = {Phys. A.},
volume = {121},
number = {3},
pages = {587-616},
year = {1983},
issn = {0378-4371},
doi = {https://doi.org/10.1016/0378-4371(83)90013-4},
url = {https://www.sciencedirect.com/science/article/pii/0378437183900134},
author = {A.O. Caldeira and A.J. Leggett},
}

@article{CALDEIRA1983374,
title = {Quantum tunnelling in a dissipative system},
journal = {Ann. Phys. (NY).},
volume = {149},
number = {2},
pages = {374-456},
year = {1983},
issn = {0003-4916},
doi = {https://doi.org/10.1016/0003-4916(83)90202-6},
url = {https://www.sciencedirect.com/science/article/pii/0003491683902026},
author = {A.O Caldeira and A.J Leggett}
}

@article{lindblad1976generators,
  title={On the generators of quantum dynamical semigroups},
  author={Lindblad, Goran},
  journal={Commun. Math. Phys.},
  volume={48},
  pages={119--130},
  year={1976},
doi={https://doi.org/10.1007/BF01608499},
  publisher={Springer}
}

@article{gorini1976completely,
  title={{Completely positive dynamical semigroups of $N$-level systems}},
  author={Gorini, Vittorio and Kossakowski, Andrzej and Sudarshan, Ennackal Chandy George},
  journal={J. Math. Phys.},
  volume={17},
  number={5},
  pages={821--825},
  year={1976},
  publisher={American Institute of Physics},
doi={https://doi.org/10.1063/1.522979}
}

@article{van1985elimination,
title = {Elimination of fast variables},
journal = {Phys. Rep.},
volume = {124},
number = {2},
pages = {69-160},
year = {1985},
issn = {0370-1573},
doi = {https://doi.org/10.1016/0370-1573(85)90002-X},
url = {https://www.sciencedirect.com/science/article/pii/037015738590002X},
author = {N.G. {Van Kampen}},

}

@article{kramers1940brownian,
title = {Brownian motion in a field of force and the diffusion model of chemical reactions},
journal = {Physica},
volume = {7},
number = {4},
pages = {284-304},
year = {1940},
issn = {0031-8914},
doi = {https://doi.org/10.1016/S0031-8914(40)90098-2},
url = {https://www.sciencedirect.com/science/article/pii/S0031891440900982},
author = {H.A. Kramers},
}

@article{aharonov1993quantum,
  title = {Quantum random walks},
  author = {Aharonov, Y. and Davidovich, L. and Zagury, N.},
  journal = {Phys. Rev. A.},
  volume = {48},
  issue = {2},
  pages = {1687--1690},
  numpages = {0},
  year = {1993},
  month = {Aug},
  publisher = {American Physical Society},
  doi = {10.1103/PhysRevA.48.1687},
  url = {https://link.aps.org/doi/10.1103/PhysRevA.48.1687}
}

@article{kempe2003quantum,
author = {J Kempe},
title = {{Quantum random walks: An introductory overview}},
journal = {Contemp. Phys.},
volume = {44},
number = {4},
pages = {307-327},
year = {2003},
publisher = {Taylor & Francis},
doi = {https://doi.org/10.1080/00107151031000110776}

}

@article{ATTAL20121545,
title = {Open quantum walks on graphs},
journal = {Phys. Lett. A.},
volume = {376},
number = {18},
pages = {1545-1548},
year = {2012},
issn = {0375-9601},
doi = {https://doi.org/10.1016/j.physleta.2012.03.040},
url = {https://www.sciencedirect.com/science/article/pii/S0375960112003453},
author = {S. Attal and F. Petruccione and I. Sinayskiy}
}

@article{attal2012open,
  title={Open quantum random walks},
  author={Attal, Stephane and Petruccione, Francesco and Sabot, Christophe and Sinayskiy, Ilya},
  journal={J. Stat. Phys.},
  volume={147},
  number={4},
  pages={832--852},
  year={2012},
doi = {https://doi.org/10.1007/s10955-012-0491-0},
  publisher={Springer}
}

@article{Sinayskiy_2012,
doi = {10.1088/0031-8949/2012/T151/014077},
url = {https://dx.doi.org/10.1088/0031-8949/2012/T151/014077},
year = {2012},
month = {nov},
publisher = {IOP Publishing},
volume = {2012},
number = {T151},
pages = {014077},
author = {Ilya Sinayskiy and Francesco Petruccione},
title = {{Properties of open quantum walks on $\mathbb{Z}$}},
journal = {Phys. Scr.}
}

@book{kraus1983states,
  title={{States, Effects, and Operations Fundamental Notions of Quantum Theory: Lectures in Mathematical Physics at the University of Texas at Austin}},
  author={Kraus, Karl and B{\"o}hm, Arno and Dollard, John D and Wootters, WH},
  year={1983},
  publisher={Springer}
}

@article{venegas2012quantum,
  title={Quantum walks: a comprehensive review},
  author={Venegas-Andraca, Salvador El{\'\i}as},
  journal={Quant. Inf. Proc.},
  volume={11},
  number={5},
  pages={1015--1106},
  year={2012},
  publisher={Springer},
  doi = {https://doi.org/10.1007/s11128-012-0432-5}
}

@article{sinayskiy2012efficiency,
  title={Efficiency of open quantum walk implementation of dissipative quantum computing algorithms},
  author={Sinayskiy, Ilya and Petruccione, Francesco},
  journal={Quant. Inf. Proc.},
  volume={11},
  pages={1301--1309},
  year={2012},
  publisher={Springer},
doi = {https://doi.org/10.1007/s11128-012-0426-3}
}

@article{sinayskiy2019open,
  title={{Open quantum walks: A mini review of the field and recent developments}},
  author={Sinayskiy, Ilya and Petruccione, Francesco},
  journal={Eur. Phys. J. Spec. Top.},
  volume={227},
  number={15-16},
  pages={1869--1883},
  year={2019},
  publisher={Springer},
doi = {https://doi.org/10.1140/epjst/e2018-800119-5}
}

@article{sinayskiy2014quantum,
  title={Quantum optical implementation of open quantum walks},
  author={Sinayskiy, Ilya and Petruccione, Francesco},
  journal={ Int. J. Quantum Inform.},
  volume={12},
  number={02},
  pages={1461010},
  year={2014},
  publisher={World Scientific},
doi = {https://doi.org/10.1142/S0219749914610103}

}

@article{sinayskiy2013microscopic,
  title={Microscopic derivation of open quantum walk on two-node graph},
  author={Sinayskiy, Ilya and Petruccione, Francesco},
  journal={Open
Syst. Inf. Dyn.},
  volume={20},
  number={03},
  pages={1340007},
  year={2013},
  publisher={World Scientific},
doi = {https://doi.org/10.1142/S1230161213400076}
}

@article{PhysRevA.92.032105,
  title = {Microscopic derivation of open quantum walks},
  author = {Sinayskiy, Ilya and Petruccione, Francesco},
  journal = {Phys. Rev. A.},
  volume = {92},
  issue = {3},
  pages = {032105},
  numpages = {11},
  year = {2015},
  month = {Sep},
  publisher = {American Physical Society},
  doi = {10.1103/PhysRevA.92.032105},
  url = {https://link.aps.org/doi/10.1103/PhysRevA.92.032105}
}

@article{PhysRevLett.102.180501,
  title = {{Universal Computation by Quantum Walk}},
  author = {Childs, Andrew M.},
  journal = {Phys. Rev. Lett.},
  volume = {102},
  issue = {18},
  pages = {180501},
  numpages = {4},
  year = {2009},
  month = {May},
  publisher = {American Physical Society},
  doi = {10.1103/PhysRevLett.102.180501},
  url = {https://link.aps.org/doi/10.1103/PhysRevLett.102.180501}
}

@article{PhysRevA.81.042330,
  title = {Universal quantum computation using the discrete-time quantum walk},
  author = {Lovett, Neil B. and Cooper, Sally and Everitt, Matthew and Trevers, Matthew and Kendon, Viv},
  journal = {Phys. Rev. A.},
  volume = {81},
  issue = {4},
  pages = {042330},
  numpages = {7},
  year = {2010},
  month = {Apr},
  publisher = {American Physical Society},
  doi = {10.1103/PhysRevA.81.042330},
  url = {https://link.aps.org/doi/10.1103/PhysRevA.81.042330}
}

@article{chawla2023multi,
  title={Multi-qubit quantum computing using discrete-time quantum walks on closed graphs},
  author={Chawla, Prateek and Singh, Shivani and Agarwal, Aman and Srinivasan, Sarvesh and Chandrashekar, CM},
  journal={Sci. Rep.},
  volume={13},
  number={1},
  pages={12078},
  year={2023},
  publisher={Nature Publishing Group UK London},
doi={https://doi.org/10.1038/s41598-023-39061-1}
}

@article{konno2013limit,
  title={Limit theorems for open quantum random walks},
  author={Konno, Norio and Yoo, Hyun Jae},
  journal={J. Stat. Phys.},
  volume={150},
  pages={299--319},
  year={2013},
  publisher={Springer},
doi={https://doi.org/10.1007/s10955-012-0668-6}
}

@article{sadowski2016central,
  title={Central limit theorem for reducible and irreducible open quantum walks},
  author={Sadowski, Przemys{\l}aw and Pawela, {\L}ukasz},
  journal={ Quant. Inf. Proc.},
  volume={15},
  pages={2725--2743},
  year={2016},
  publisher={Springer},
doi={https://doi.org/10.1007/s11128-016-1314-z}
}

@inproceedings{attal2015central,
  title={Central limit theorems for open quantum random walks and quantum measurement records},
  author={Attal, St{\'e}phane and Guillotin-Plantard, Nadine and Sabot, Christophe},
  booktitle={	Ann. Henri Poincaré},
  volume={16},
  pages={15--43},
  year={2015},
  organization={Springer},
doi={https://doi.org/10.1007/s00023-014-0319-3}
}

@article{pellegrini2014continuous,
  title={{Continuous time open quantum random walks and non-Markovian Lindblad master equations}},
  author={Pellegrini, Cl{\'e}ment},
  journal={J. Stat. Phys.},
  volume={154},
  number={3},
  pages={838--865},
  year={2014},
  publisher={Springer},
doi={https://doi.org/10.1007/s10955-013-0910-x}
}

@article{sinayskiy2015microscopicbrown,
  title={{Microscopic derivation of open quantum Brownian motion: a particular example}},
  author={Sinayskiy, Ilya and Petruccione, Francesco},
  journal={Phys. Scr.},
  volume={2015},
  number={T165},
  pages={014017},
  year={2015},
doi={http://dx.doi.org/10.1088/0031-8949/2015/T165/014017},
  publisher={IOP Publishing}
}

@article{sinayskiy2017steady,
  title={{Steady-State control of open Quantum Brownian Motion}},
  author={Sinayskiy, Ilya and Petruccione, Francesco},
  journal={Fortschr. Phys.},
  volume={65},
  number={6-8},
  pages={1600063},
  year={2017},
  publisher={Wiley Online Library},
doi={https://doi.org/10.1002/prop.201600063}
}

@article{PhysRevA.107.062206,
  title= {{Hybrid completely positive Markovian quantum-classical dynamics}},
  author= {Di\'osi, Lajos},
  journal= {Phys. Rev. A.},
  volume= {107},
  issue= {6},
  pages= {062206},
  numpages= {10},
  year= {2023},
  month= {Jun},
  publisher= {American Physical Society},
  doi= {10.1103/PhysRevA.107.062206},
  url= {https://link.aps.org/doi/10.1103/PhysRevA.107.062206}
}

@article{L.Diósi_1993,
doi = {10.1209/0295-5075/22/1/001},
url = {https://dx.doi.org/10.1209/0295-5075/22/1/001},
year = {1993},
month = {apr},
publisher = {},
volume = {22},
number = {1},
pages = {1},
author = {L. Diósi},
title = {{On High-Temperature Markovian Equation for Quantum Brownian Motion}},
journal = {EPL.},
}

@article{homa2019positivity,
  title={{Positivity violations of the density operator in the Caldeira-Leggett master equation}},
  author={Homa, G{\'a}bor and Bern{\'a}d, J{\'o}zsef Zsolt and Lisztes, L{\'a}szl{\'o}},
  journal={Eur. Phys. J. D.},
  volume={73},
  pages={1--13},
  year={2019},
  publisher={Springer},
doi={https://doi.org/10.1140/epjd/e2019-90604-4}
}

@book{carmichael1999statistical,
  title={{Statistical Methods in Quantum Optics 1: Master Equations and Fokker-Planck Equations}},
  author={Carmichael, Howard},
  volume={1},
  year={2002},
  publisher={Berlin: Springer}
}

@book{scully1999quantum,
  title={Quantum optics},
  author={Scully, Marlan O and Zubairy, M Suhail},
  year={1997},
  publisher={Cambridge: Cambridge University Press}
}

@article{layton2024healthier,
  title={A healthier semi-classical dynamics},
  author={Layton, Isaac and Oppenheim, Jonathan and Weller-Davies, Zachary},
  journal={Quantum},
  volume={8},
  pages={1565},
  year={2024},
  publisher={Verein zur F{\"o}rderung des Open Access Publizierens in den Quantenwissenschaften},
doi={https://doi.org/10.22331/q-2024-12-16-1565}
}

@article{layton2024classical,
  title={The classical-quantum limit},
  author={Layton, Isaac and Oppenheim, Jonathan},
  journal={PRX Quantum},
  volume={5},
  number={2},
  pages={020331},
  year={2024},
  publisher={APS},
doi={https://doi.org/10.1103/PRXQuantum.5.020331}
}

@article{oppenheim2023postquantum,
  title={A postquantum theory of classical gravity?},
  author={Oppenheim, Jonathan},
  journal={Phys. Rev. X.},
  volume={13},
  number={4},
  pages={041040},
  year={2023},
  publisher={APS},
doi={https://doi.org/10.1103/PhysRevX.13.041040}
}

@article{oppenheim2022two,
  title={The two classes of hybrid classical-quantum dynamics},
  author={Oppenheim, Jonathan and Sparaciari, Carlo and {\v{S}}oda, Barbara and Weller-Davies, Zachary},
  journal={arXiv:2203.01332},
  year={2022},
url={https://doi.org/10.48550/arXiv.2203.01332}
}

@article{halliwell1998effective,
  title={{Effective theories of coupled classical and quantum variables from decoherent histories: A New approach to the back reaction problem}},
  author={Halliwell, JJ},
  journal={Phys. Rev. D.},
  volume={57},
  number={4},
  pages={2337},
  year={1998},
  publisher={APS},
doi={https://doi.org/10.1103/PhysRevD.57.2337}
}

@article{tilloy2024general,
  title={General quantum-classical dynamics as measurement based feedback},
  author={Tilloy, Antoine},
  journal={SciPost Physics},
  volume={17},
  number={3},
  pages={083},
  year={2024},
doi={ 10.21468/SciPostPhys.17.3.083}
}

@article{Gardiner1984,
  title = {{Adiabatic elimination in stochastic systems. I. Formulation of methods and application to few-variable systems}},
  author = {Gardiner, C. W.},
  journal = {Phys. Rev. A.},
  volume = {29},
  issue = {5},
  pages = {2814--2822},
  numpages = {0},
  year = {1984},
  month = {May},
  publisher = {American Physical Society},
  doi = {10.1103/PhysRevA.29.2814},
  url = {https://link.aps.org/doi/10.1103/PhysRevA.29.2814}
}

@article{Ankerhold2001,
  title = {{Strong Friction Limit in Quantum Mechanics: The Quantum Smoluchowski Equation}},
  author = {Ankerhold, Joachim and Pechukas, Philip and Grabert, Hermann},
  journal = {Phys. Rev. Lett.},
  volume = {87},
  issue = {8},
  pages = {086802},
  numpages = {4},
  year = {2001},
  month = {Aug},
  publisher = {American Physical Society},
  doi = {10.1103/PhysRevLett.87.086802},
  url = {https://link.aps.org/doi/10.1103/PhysRevLett.87.086802}
}

@book{schleich2001,
  title={{Quantum Optics in Phase Space}},
  author={Schleich, Wolfgang P.},
  year={2001},
  publisher={Wiley-VCH Verlag Berlin GmbH, Berlin}
}

@article{SKINNER1979561,
title = {{Derivation of Smoluchowski equations with corrections for Fokker-Planck and BGK collision models}},
journal = {{Physica A: Stat. Mech. Appl.}},
volume = {96},
number = {3},
pages = {561-572},
year = {1979},
issn = {0378-4371},
doi = {https://doi.org/10.1016/0378-4371(79)90013-X},
url = {https://www.sciencedirect.com/science/article/pii/037843717990013X},
author = {James L. Skinner and Peter G. Wolynes}
}

@article{Diósi_2014,
doi = {10.1088/0031-8949/2014/T163/014004},
url = {https://doi.org/10.1088/0031-8949/2014/T163/014004},
year = {2014},
month = {dec},
publisher = {IOP Publishing},
volume = {2014},
number = {T163},
pages = {014004},
author = {Diósi, Lajos},
title = {Hybrid quantum-classical master equations},
journal = {Physica Scripta}
}

@article{zungu2026microscopic,
  title={{Microscopic derivation of a completely positive master equation for the description of Open Quantum Brownian Motion of a particle in a potential}},
  author={Zungu, Ayanda and Sinayskiy, Ilya and Petruccione, Francesco},
  journal={arXiv:2602.03534}, year={2026},
url ={https://doi.org/10.48550/arXiv.2602.03534}
}

@article{Carlos2025,
doi = {10.1088/1751-8121/ae21a7},
url = {https://doi.org/10.1088/1751-8121/ae21a7},
year = {2025},
month = {nov},
publisher = {IOP Publishing},
volume = {58},
number = {48},
pages = {485303},
author = {de la Iglesia, Manuel D and Lardizabal, Carlos F},
title = {{Exact solutions of open quantum Brownian motions on the real line for two-level systems}},
journal = {Journal of Physics A: Mathematical and Theoretical}
}

@article{PhysRevE.81.021107,
  title = {{Quantum Smoluchowski equation: A systematic study}},
  author = {Maier, Stefan A. and Ankerhold, Joachim},
  journal = {Phys. Rev. E},
  volume = {81},
  issue = {2},
  pages = {021107},
  numpages = {14},
  year = {2010},
  month = {Feb},
  publisher = {American Physical Society},
  doi = {10.1103/PhysRevE.81.021107},
  url = {https://link.aps.org/doi/10.1103/PhysRevE.81.021107}
}

@article{PhysRevA.94.042123,
  title = {{Lindblad model of quantum Brownian motion}},
  author = {Lampo, Aniello and Lim, Soon Hoe and Wehr, Jan and Massignan, Pietro and Lewenstein, Maciej},
  journal = {Phys. Rev. A.},
  volume = {94},
  issue = {4},
  pages = {042123},
  numpages = {13},
  year = {2016},
  month = {Oct},
  publisher = {American Physical Society},
  doi = {10.1103/PhysRevA.94.042123},
  url = {https://link.aps.org/doi/10.1103/PhysRevA.94.042123}
}

@article{PhysRevLett.84.1374,
  title = {{Completely Positive Quantum Dissipation}},
  author = {Vacchini, Bassano},
  journal = {Phys. Rev. Lett.},
  volume = {84},
  issue = {7},
  pages = {1374--1377},
  numpages = {0},
  year = {2000},
  month = {Feb},
  publisher = {American Physical Society},
  doi = {10.1103/PhysRevLett.84.1374},
  url = {https://link.aps.org/doi/10.1103/PhysRevLett.84.1374}
}
\end{document}